\documentclass{jfm}
\usepackage[utf8]{inputenc}
\usepackage[english]{babel}
\usepackage{graphicx}
\usepackage{subcaption}
\usepackage[font=small, labelfont=sc, labelsep=period]{caption}
\usepackage{amsmath}
\usepackage[colorlinks=true,citecolor=blue,linkcolor=blue]{hyperref}
\usepackage{epstopdf, epsfig}
\newcommand{\mean}[1]{\left<#1\right>}
\newcommand{\myvector}[1]{\mathbf{#1}}
\newcommand{\dif}{\mathrm{d}}
\newcommand{\norm}[1]{\|#1\|}

\shorttitle{Interaction between hairy surfaces and turbulence}
\shortauthor{J. Sundin and S. Bagheri}

\title{Interaction between hairy surfaces and turbulence for different surface time scales}
\author{Johan Sundin\aff{1} \and Shervin Bagheri\aff{1} \corresp{\email{shervin@mech.kth.se}}}
\affiliation{\aff{1}Linné FLOW Centre, KTH Mechanics, Royal Institute of Technology, SE-100 44 Stockholm,
Sweden}

\begin{document}

\maketitle
\begin{abstract}
Surfaces with filamentous structures are ubiquitous in nature on many different scales, ranging from forests to micrometer-sized cilia in organs. Hairy surfaces are elastic and porous, and it is not fully understood how they modify turbulence near a wall. The interaction between hairy surfaces and turbulent flows is here investigated numerically in a turbulent channel flow configuration at $\Rey_\tau \approx 180$. We show that a filamentous bed of a given geometry can modify a turbulent flow very differently depending on the resonance frequency of the surface, which is determined by elasticity and mass of the filaments.
Filaments having resonance frequencies lower than the main frequency content of the turbulent wall-shear stress  conform to slowly traveling  elongated streaky structures, since they are too slow to adapt to fluid forces of higher frequencies. On the other hand, a bed consisting of stiff and low-mass filaments has a high resonance frequency and shows local regions of increased permeability, which results in large entrainment
and a vast increase in drag.

\end{abstract}


\section{Introduction}
Surfaces found in nature often have deformable filamentous surface textures. At atmospheric scales,
the understanding of turbulence over aerial and aquatic vegetation is of great importance for ecological, environmental and industrial applications. For example, an optimal placement of  wind turbines  \citep{hansen15} requires relatively accurate wind predictions, which in turn is determined by interaction
of a turbulent boundary layer over  different terrains, such as forests. In ecosystems, the interaction between a boundary layer and a bed of seagrass is essential for controlling the provision of nutrients to the plants, the scattering of pollen, etc.~\citep{nepf12}.

At smaller scales, turbulent flows over filamentous structures are observed around and inside organisms. The fur of seals have been found to form riblet-like grooves, resulting in drag reduction \citep{itoh06}. Filament-like flow sensors are used by fish and flying insects, serving as inspiration for artificial sensors \citep{tao12}. Fish have superficial neuromasts and neuromasts contained in channels on their sides, termed the lateral line, enabling them to sense the velocity field as well as the pressure distribution along the body. The lateral line, in particular, has inspired artificial underwater-sensing technology, termed artificial lateral lines \citep{liu16}.

The dynamics of a hairy surface is characterised by a certain time scale, because the speed of the filaments is limited by their inertia or, in some cases, by the viscous damping. When inertia dominates over viscous damping, the characteristic time scale found from the Euler-Bernoulli equation is
\begin{equation}
    T \sim l^2\sqrt{\frac{\rho_\mathrm{s} A + \chi}{EI}}.
    \label{eq:order_of_magnitude_time}
\end{equation}
Here, $l$ is the length, $\rho_\mathrm{s}$ the density, $A$ the cross-sectional area, $E$ the Young's modulus and $I$ the area moment of inertia of a filament. The  constant $\chi$ represents the added mass. The bed of filaments is a porous medium of finite permeability as well as an elastic medium which can deform, thus making it an anisotropic poroelastic medium.
The objective of this paper is to show the effects of a filamentous bed on turbulence for different surface time scales $T$,
using direct numerical simulations (DNS).

In particular, we want to characterise the two-way coupling between the surface and the turbulence through a time scale analysis. In general, the temporal behavior of turbulence is of broadband character, but the frequency-weighted spectrum of wall shear-stress has a peak value for a range of Reynolds numbers \citep{hu06}. This suggests that one may associate a characteristic dominant time scale $T_\mathrm{f}$ with the turbulent flow near a wall. In simplified settings \citep{jimenez91}, this time scale can also be related to cyclic turbulent events involving near-wall quasi-streamwise vortices.
%
 %
 As we will show in this paper, the response of the bed to the forcing induced by the wall turbulence and the modification of the turbulent flow due to the movement of the surface are dramatically different for $T\ll T_\mathrm{f}$ and $T\gg T_\mathrm{f}$.


In this investigation filaments are attached to one channel wall. They are placed densely enough to create a strong coupling between adjacent filaments, but with a distance large enough so that they rarely touch each other. This makes it possible to resolve each individual filament. We use one fixed filament geometry, whereas the mass density and the elasticity of the filaments are varied, changing the time scale of the bed. To the best of the authors knowledge, there are no earlier numerical investigations of the interaction between an anisotropic poroelastic medium and turbulence where the  microstructure of the bed is fully resolved.

The work that most closely resembles this study is the experimental investigation by \citet{brucker11}. He characterised the interaction between filamentous beds -- which were larger and more sparse compared to our configuration -- with near-wall turbulence in an oil channel. The pillars were fabricated using PDMS. For a specific non-uniform filament arrangement in the streamwise and spanwise directions, \citet{brucker11} reported a stabilisation of streamwise streaks, and proposed that such beds could be used to reduce drag, although this remains to be shown.
There has been substantial work on turbulent flows over vegetation, which have similarities to our study. \citet{nepf12} provides an excellent review of how canopy-scale fluid instabilities and waves modify the transfer of mass and moment between the free flowing fluid and the bed for aquatic vegetation. \citet{delangre08} reviews the effect of wind over canopies, showing that the reduced velocity and the Cauchy number -- which characterize dynamical effects and mean filament displacement, respectively -- need to be $\mathcal{O}(1)$ for a strong interaction between the wind and the canopy.
In different contexts than wall-bounded turbulence, a number of previous studies  on flows over hairy surfaces demonstrate a strong interaction between slender structures and flows when certain spatial and temporal scales are matched; examples include the analysis of surfaces covered in carbon nanotubes \citep{battiato10}, the study of how plants reconfigure to reduce drag \citep{gosselin11} and the study of flow past a cylinder with a hairy coating \citep{favier09}.

Finally, there exists extensive previous numerical work on turbulent flows over porous media \citep{jimenez01, breugem06, rosti18} as well as over compliant surfaces \citep{kim14, rosti17}. The prominent effect of porous walls is -- similar to canopy flows -- significant increase in both drag and entrainment  induced by large-scale spanwise vortices. System-size instabilities are also often observed of flows over compliant walls, related to large-amplitude quasi-two-dimensional traveling surface waves. In contrast to this work, all these previous efforts consider porosity and elasticity separate from each other, and thus are not able to connect a characteristic time scale  to a specific physical  geometry of the bed. As we will show,  significant increase in drag and entrainment can also be rooted in intrinsic microscopic surface properties, not necessarily induced by macroscopic instabilities.


We characterise surfaces where the density ratio between the filaments and the fluid is in the range $1$ to $1000$. This is motivated by the fact that there are many materials with a density similar to water, such as organic materials and plastics, however, few materials are lighter than air. Hence, filament beds in water, such as aquatic vegetation, tend to have a density similar to the surrounding fluid, while filament beds in air, such as a forest, tend to be much heavier than the surrounding fluid.

Numerically, the fluid flow is described by a lattice-Boltzmann method and the interaction with the filaments by an immersed-boundary method. Filament dynamics is described by a discretisation of the Euler-Bernoulli equation, where inertia is taken into account. The flow has a friction Reynolds number of $Re_\tau = h u_\tau/\nu \approx 180$, with channel half-height $h$, kinematic viscosity $\nu$ and friction velocity $u_\tau = \sqrt{\tau_\mathrm{wall}/\rho}$, where $\tau_\mathrm{wall}$ is the effective total shear stress at the wall of interest. To drive the flow, a constant pressure gradient is used, giving $Re_\tau = 180$ for a symmetric smooth channel.


The rest of the article is structured as follows. In section \ref{sec:transfer_functions}, we make an order-of-magnitude approximation of the filament time scale, resulting in eq.~\eqref{eq:order_of_magnitude_time} and also discuss the turbulent time scales. The numerical method
is described in section \ref{sec:numerics}. In section \ref{sec:results}, we show that the movement of heavy (and thus slow) filaments have a negligible impact on turbulence, whereas for lighter (and thus faster) filaments the turbulent wall-shear stress  induce a high local permeability, which in turn increases the drag and the isotropy of the velocity field, as well as the entrainment into the bed. In section \ref{sec:extendend_discussion}, we present a simple fluid-structure interaction model and compare it to numerical simulations of filamentous beds with different characteristics.   Finally, conclusions are provided in section \ref{sec:conclusion}.

\section{Characterisation of time scales}
In this section, we discuss the fluid forces on the filaments and provide  order-of-magnitude estimates of the filament time scale (section \ref{sec:order_of_magnitude_estimation}) and  turbulence time scales (section \ref{sec:comparision_to_turbulence}). Lastly, the configurations that will be investigated computationally are briefly presented in section \ref{sec:simulation_parameters}.


\label{sec:transfer_functions}
\begin{figure}
  \begin{subfigure}{0.45\textwidth}
      \centering
      \includegraphics[width=8cm]{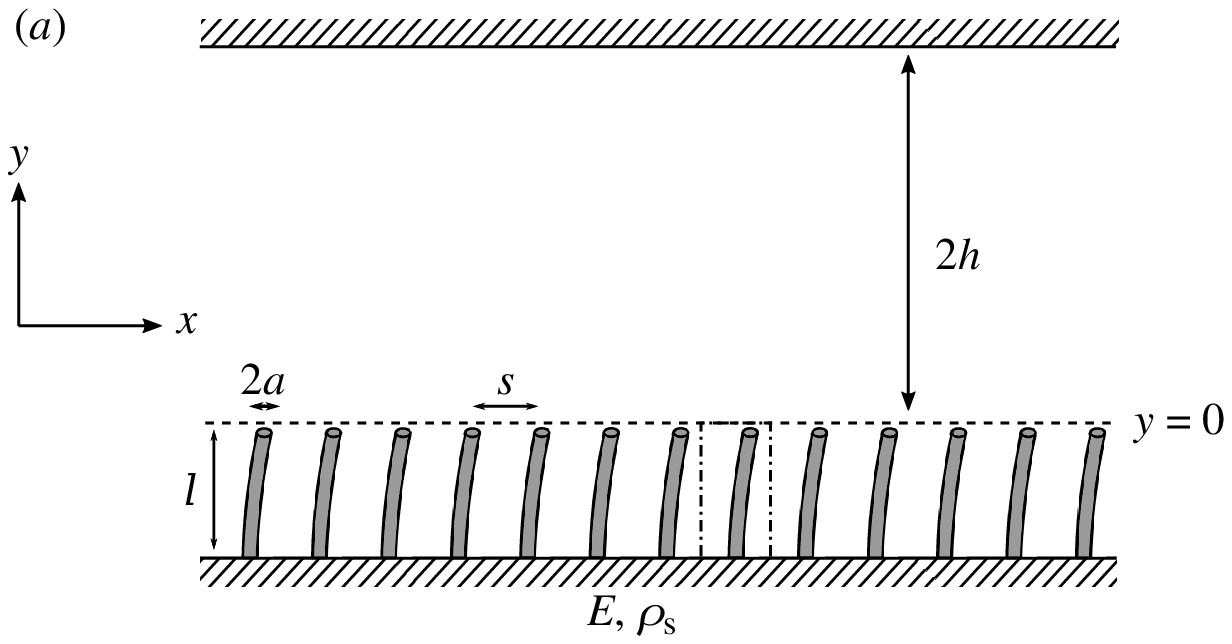}
  \end{subfigure}
  \begin{subfigure}{0.45\textwidth}
      \centering
      \hspace{2cm}
      \includegraphics[width=3.3cm]{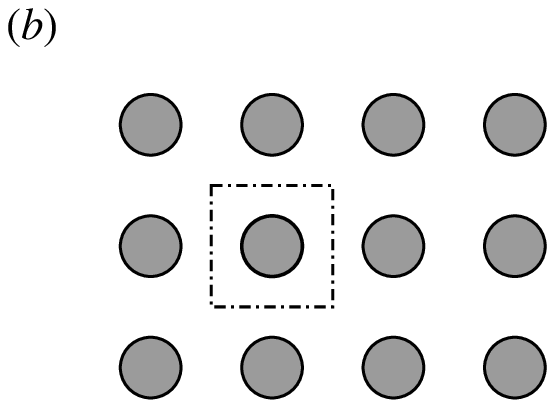}
  \end{subfigure}
  \caption{Illustration of the geometry and the filament parameters: ($a$) side view and ($b$) top view. Geometrical parameters are given in ($a$). The dashed line in ($a$) shows the location of $y=0$ (at the tip of the filaments) and the dash-dotted rectangle marks one cell. A square packing structure, as illustrated in ($b$), is used in the simulations.}
  \label{fig:geometry}
\end{figure}

A schematic of the filament geometry is shown in fig.~\ref{fig:geometry}, with a side view in \ref{fig:geometry}$a$ and a top view in \ref{fig:geometry}$b$. The filaments with density $\rho_\mathrm{s}$, Young's modulus $E$ and length $l$ are assumed to have a circular cross section with radius $a$. We assume the resting position of the filaments to be straight vertically, packed in a square lattice structure. The center-to-center distance is denoted by $s$.  

The force on a filament can be divided into two contributions. The first is due to the three-dimensional effects at the tip of the filaments, the tip force, $F_\mathrm{tip}$, and the second is due to the drag distributed along the body of a filament, $f_\mathrm{body}$. For simplicity, they can be treated as independent of each other. These forces give rise to a movement of the filaments, described by the Euler-Bernoulli equation,
\begin{equation}
    EI\frac{\partial^4 q}{\partial y^4} + (\rho_\mathrm{s} A + \chi)\frac{\partial^2 q}{\partial t^2} = f_\mathrm{body}.
    \label{eq:Euler-Bernoulli}
\end{equation}
Here, $q = q(y, t)$ is the streamwise displacement, with $y$ being the wall-normal direction and $t$ the time, and $I = \frac{\pi}{4}a^4$ is the area moment of inertia corresponding to that of a cylinder. The first term represents the force due to the deflection of the filament, whereas the second describes the inertial force of the acceleration.
The constant $\chi$ accounts for the added mass.
The boundary conditions are
\begin{equation}
\begin{array}{lcll}
    q = 0 &\text{ and } &\dfrac{\partial q}{\partial y} = 0 &\text{ at the base } (y = -l),\\
    \dfrac{\partial^2 q}{\partial y^2} = 0 &\text{ and } &EI\dfrac{\partial^3 q}{\partial y^3} = -F_\mathrm{tip} &\text{ at the tip } (y = 0).
\end{array}
\label{eq:boundary_conditions}
\end{equation}
The conditions at the base correspond to clamped beam, while the conditions at the tip correspond to zero applied torque and an applied tip force. The filaments are attached to the surface at $y = -l$ and the tips are located at approximately $y = 0$, as shown in fig.~\ref{fig:geometry}. Equation (\ref{eq:Euler-Bernoulli}) holds under the assumption of small displacements and zero axial tension. The flexural rigidity $B = EI$ is the  key parameter for characterising the bending of filaments, and thus allows one to consider more general filament geometries than considered here. However, for simplicity -- and to be coherent with the filaments used in our DNS -- we assume that the filaments are homogeneous with $I= \frac{\pi}{4}a^4$. We also neglect internal damping due to dissipation in the filament material.


When a filament accelerates, it displaces fluid, and the extra force needed to accelerate the fluid can be incorporated through the added mass, $\chi$.  The added mass is negligible when $\rho_\mathrm{s} \gg \rho$, but is significant for lighter filaments.  For the filament bed the added mass can be used as a crude model of the filament-filament coupling. If adjacent filaments move in phase  , the fluid of a cell, illustrated in fig.~\ref{fig:geometry}, can be assumed to move with the filament, so that the added mass can be modelled as
\begin{equation}
    \chi = \rho (s^2 - A).
\label{eq:added_mass}
\end{equation}
This approximation can physically be motivated for beds where the center-to-center distance is comparable to the diameter of the filaments. For the beds considered here, $s^2 = 16a^2>A$, and we make a further approximation, namely, that $\chi \approx \rho s^2$.
\subsection{Non-dimensional filament parameters}

\label{sec:order_of_magnitude_estimation}
If the filaments are placed densely, the mean fluid velocity inside the bed is very small. The slow flow inside bed means that the force on the filament tips is much larger than the force on the body of the filaments, $F_\mathrm{tip} \gg f_\mathrm{body}l$. This estimation is discussed more quantitatively in appendix \ref{sec:transfer_function_derivation}. The wall-shear stress of the actual wall, to which the filaments are attached, is negligible.
The filament tip force in the streamwise direction can therefore be approximated by the  fluid shear stress, $\tau$, on the top face of the cell coinciding with the $y=0$ plane (see fig.~\ref{fig:geometry}),
\[
F_\mathrm{tip}
\approx \int_\mathrm{cell~face}\tau \dif x \dif z.
\]
Next, we non-dimensionalise the filament equation by defining $q^*=q/a$, $y^*=y/l$, $t^*=t/T_\mathrm{f}$ and $F^*_{\mathrm{tip}}=F_{\mathrm{tip}}/\langle F_{\mathrm{tip}}\rangle$. The filament displacement is thus assumed to scale with the radius, $a$, whereas the characteristic length scale in the wall-normal direction is $l$. We have also introduced a reference time scale $T_\mathrm{f}$ and a characteristic tip force magnitude $\langle F_{\mathrm{tip}}\rangle$. Henceforth, $\mean{\cdot}$ will be used to denote the mean value of a quantity. In dimensionless variables, we have
\begin{equation}
    \frac{\partial^4 q^*}{\partial {y^*}^4} + {T^*}^2\frac{\partial^2 q^*}{\partial {t^*}^2} = 0
    \label{eq:Euler-Bernoulli:non-dim}
\end{equation}
and $q^* = 0$, ${\partial q^*}/{\partial y^*} = 0$ at $y^*=-1$ and ${\partial^2 q^*}/{\partial {y^*}^2}=0$, ${\partial^3 q^*}/{\partial {y^*}^3} = -Q^* F^*_\mathrm{tip}$ at $y^*=0$.
We obtain two non-dimensional numbers, namely,
\begin{equation}
    T^*=\frac{T}{T_\mathrm{f}} = \frac{1}{T_\mathrm{f}}\left (l^2\sqrt{\frac{\rho_\mathrm{s} A + \chi}{EI}}\right)
\end{equation}
and
\begin{equation}
    Q^*=\frac{Q}{a} = \frac{\langle F_{\mathrm{tip}}\rangle l^3}{EI a}.
    \label{eq:cauchy_number}
\end{equation}
The latter number represents a Cauchy number, which describes the static deformation under the effect of shear force, i.e. $Q^*\sim \mean{q^*(0,t)}$,  where $\mean{q^*(0,t)}$ is the non-dimensional mean displacement of the filament tip. Therefore, we may expect a bending of the filament under an external tip force $\mean{F_\mathrm{tip}}$ when the filament is sufficiently long and soft so that $Q^*\sim 1$. The displacement of the tip is henceforth denoted with a tilde, so that $\tilde{q}(t) = q(0,t)$.
%

The second non-dimensional number  $T^*$ is commonly referred to as the reduced velocity.  The numerator $T$ is proportional  the period of natural free vibrations of a filament,
\begin{equation}
    T_\mathrm{n} = \alpha l^2\sqrt{\frac{A\rho_\mathrm{s} + \chi}{EI}}.
    \label{eq:exact_natural_frequency}
\end{equation}
where $\alpha = 2\pi/1.875^2$.
Therefore, sufficiently light and stiff filaments with $T<T_\mathrm{f}$ are quick to adapt to external changes on the time scale $T_\mathrm{f}$. On the other hand,  very  heavy and soft filaments with $T>T_\mathrm{f}$ are expected to adapt comparatively slow. A strong fluid-structure interaction is thus expected when both $T^*$ and $Q^*$ are order one.

\begin{figure}
    \centering
    \begin{subfigure}{0.45\textwidth}
        \centering
        \includegraphics[width=5.5cm]{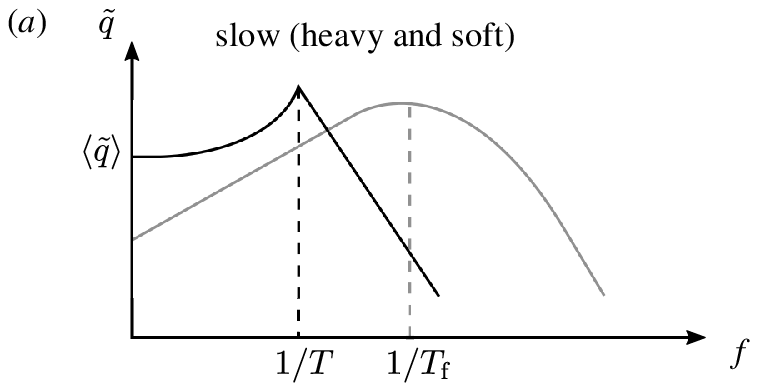}
    \end{subfigure}
    \begin{subfigure}{0.45\textwidth}
        \centering
        \includegraphics[width=5.5cm]{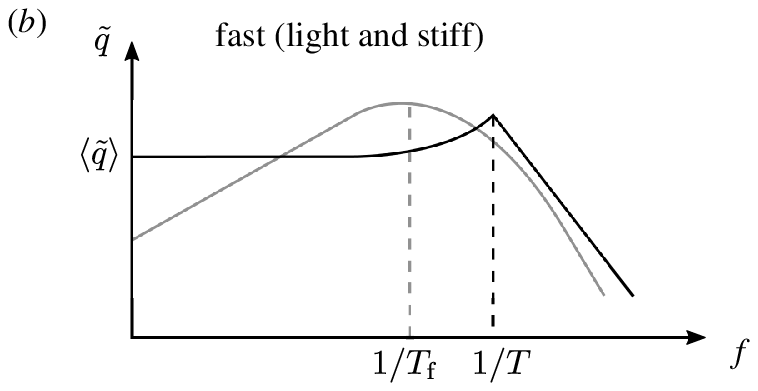}
    \end{subfigure}
    \caption{Black solid lines represent a schematic of the transfer function between the forcing and the displacement of heavy/soft filaments ($a$) and light/stiff filaments ($b$). The natural frequency is approximately $1/T$, and the mean displacement is $\mean{\tilde{q}}$. The gray line represents the frequency-weighted wall-shear stress spectrum of a turbulent channel flow, which reaches its peak value at $1/T_\mathrm{f}$.}
    \label{fig:schematic_transfer_function}
\end{figure}

\subsection{Estimates of turbulent scales}
\label{sec:comparision_to_turbulence}
In order to determine $Q^*$ and $T^*$ for a given filamentous bed, we need to estimate the mean force $\mean{F_\mathrm{tip}}$ and a characteristic forcing time scale $T_\mathrm{f}$ for a turbulent flow. The former can be estimated from the wall shear stress,
\[
    \mean{F_\mathrm{tip}}=\tau_\mathrm{wall}s^2=\rho\frac{Re_\tau^2 \nu^2}{h^2} s^2.
\]

The forcing time scale $T_\mathrm{f}$, related to the force imposed on the filaments from the flowing fluid ($F_\mathrm{tip}$), can be estimated from numerical simulations.
%
%
The frequency content of the streamwise and spanwise wall-shear stress for turbulent channel flows with smooth walls are of broadband character. However, there is a range of frequencies that dominate and which are related to the passing of  turbulent structures (e.g. streaks, vortices).
\citet{hu06} computed frequency-weighted wall-shear stress spectra for $\Rey_\tau = 180$ (and higher Reynolds numbers). Their spectra represent the energy content of the wall-shear stress for different frequencies.  A sketch of one spectrum is shown in fig.~\ref{fig:schematic_transfer_function} (gray line). For low frequencies, the frequency-weighted wall-shear stress increases almost linearly, with a derivative of one in wall-units, up to a peak. The peaks are found at $f^+ = 0.012$ and $f^+ = 0.037$, for the streamwise and spanwise components respectively. Here, $f^+=\nu/(T_\mathrm{f} u_\tau^2)$. Due to the linear increase, the energy is practically negligible one decade below the peak. The magnitude of frequencies larger than the peak  decrease rapidly, and the energy above $f^+ = 0.2$ is very small, for both components.

The forcing on a filament bed is thus dominated by frequencies around $f^+ = 0.01$, and different fluid-surface interaction behaviour can be expected, depending on the reduced velocity $T^* = T/T_\mathrm{f}$, as schematically shown in fig.~\ref{fig:schematic_transfer_function}. If the time scale of the filaments is much larger than $T_\mathrm{f}$, i.e.
$T \gg T_\mathrm{f}$ (fig.~\ref{fig:schematic_transfer_function}a), the filaments have no time to respond to the forcing and thus behave as rigid. At the other extreme, a bed of filaments with $T \ll T_\mathrm{f}$ (fig.~\ref{fig:schematic_transfer_function}b) will quickly adapt to the forcing and thus equilibrate.

We will in section \ref{sec:results} look at two surfaces whose time scales $T^*$ differ nearly by an order of magnitude, but whose expected filament displacements are of the same order of magnitude, i.e. $\mean{\tilde{q}}/a =0.64$. This is illustrated schematically in fig.~\ref{fig:schematic_transfer_function}. In this work, we assume that the bed geometry (set by $a$, $l$ and $s$) is fixed;  therefore and as apparent from eq.~\eqref{eq:order_of_magnitude_time} the expected mean deflection, $Q^*$ -- determined by $E$ -- and the expected time scale ratio, $T^*$ -- determined by both $E$ and $\rho_\mathrm{s}$ -- can be chosen independently.

\subsection{Investigated configurations}
\label{sec:simulation_parameters}

\begin{table}
    \centering
    \begin{tabular}{lllllll}
        Case\hspace{0.5cm} & Traits & $s/a$ & $\mean{\tilde{q}}/a$ & $\rho_\mathrm{s}/\rho$ & $f_\mathrm{n}^+$ & $T_\mathrm{n}/T_\mathrm{f}$\\  [3pt] 
        A & Rigid  & 4     & ---   & --- & ---  & --- \\
        B & Rigid, permeable & 8     & ---   & --- & --- & ---\\
        I & Flexible, slow & 4 & 0.64 & 300 & 0.0080 & 1.5 \\
        II & Flexible, resonant & 4 & 0.64 & 30  & 0.024 & 0.51 \\
        III & Flexible, fast & 4 & 0.64 & 1   & 0.057 & 0.21 \\
        IV & Flexible, slow, small displ & 4 & 0.24 & 800 & 0.0081 & 1.5 \\
        V & Flexible, resonant, small displ \hspace{0.5cm} & 4 & 0.24 & 85  & 0.024 & 0.50  \\
        VI & Flexible, fast, small displ & 4 & 0.24 & 11  & 0.057 & 0.21 \\
    \end{tabular}
    \caption{Geometric and dynamic properties of the configurations that are investigated. For all configurations, the radius is $a^+ = 2$ and the aspect ratio is $l/a = 10$. The non-dimensional numbers $Q^* = 3\mean{\tilde{q}}/a$ and $T^* = \alpha^{-1} T_\mathrm{n}/T_\mathrm{f}$, where $\alpha = 2\pi/1.875^2 \approx 1.8$. The mean displacements $\mean{\tilde{q}}$ are calculated from eq.~\eqref{eq:mean_disp} and the frequencies $f_\mathrm{n}^+$ from \eqref{eq:exact_natural_frequency}. Two of the cases have rigid filaments, namely A and B. For the cases with flexible filaments, I--VI, two mean displacement amplitudes are considered and for each mean displacement, three different resonance frequencies are investigated. For the quantities scaled in wall-units and for estimation of the wall-shear stress, the friction Reynolds number $\Rey_\tau = 180$ is used. The time scale ratio (reduced velocity) considers the natural frequency of the filaments and the frequency of the maximum frequency-weighted streamwise wall-shear stress.}
    \label{tab:filament_cases}
\end{table}
To characterise the interaction between turbulence and a filament bed for different filament time scales, we perform a number of simulations (tab.~\ref{tab:filament_cases}). For all simulations, a fixed filament radius is used, which in wall-units corresponds to $a^+ \approx 2$, with exact equality when $\Rey_\tau = 180$. The aspect ratio of the filaments we set to $l/a = 10$. Two cases of rigid filaments are investigated: one with a center-to-center distance of $s/a = 4$ and one with $s/a = 8$, denoted as  A and B, respectively. In addition to these, six flexible filaments configurations are investigated (I--VI), with two different mean displacements, $\mean{\tilde{q}}$, and three different natural frequencies, $f_\mathrm{n}$. 

\section{Numerical method}
\label{sec:numerics}
This section describes the numerical method for the description of the flowing fluid (section \ref{sec:fluid_solver}) and the fluid-filament interaction (section \ref{sec:solid_solver}).

\subsection{Fluid solver}
\label{sec:fluid_solver}
The lattice-Boltzmann method (LBM) is a discretisation of the Boltzmann kinetic equation. However, only the necessary details of molecular motion are retained in order to recover macroscopic conservation laws of mass, momentum and energy \citep{kruger17}.
%
%
\begin{table}
  \centering
  \begin{tabular}{lllll}
    Grid & Bulk           & $40 \geq y^+ > 10$      & $10 \geq y^+$ & Lagrangian grid   \\  [3pt]
    G1   & $\delta x^+ = 2$ & $\delta x^+ = 1$   & $\delta x^+ = 1$ & $N_h = 20$, $N_n = 16$, $N_n^{\text{lid}} = 32$ \\
    G2   & $\delta x^+ = 3$ & $\delta x^+ = 1.5$ & $\delta x^+ = 1.5$ & ---\\
    G3   & $\delta x^+ = 3$ & $\delta x^+ = 1.5$ &
      \begin{tabular}[l]{@{}ll}  $\delta x^+ = 0.75$ & at lower wall\\$\delta x^+ = 1.5$ & at upper wall\end{tabular} & $N_h = 27$, $N_n = 20$, $N_n^{\text{lid}} = 45$ \\
  \end{tabular}
  \caption{Grids used for validation and in simulations. The resolution of different grid refinements together with the properties of the Lagrangian grid of the filaments are specified.   G1 and G2 have one grid refinement at each wall, whereas G3 has one additional refinement at the lower wall. Friction Reynolds number $\Rey_\tau = 180$ is used for scaling in wall units.}
  \label{tab:grids}
\end{table}
%
In the LBM, the spatial dimensions are discretised on a grid and the particle velocity space into a set of discrete velocities. With $d$ spatial dimensions and the velocity set having a size $q$, the velocity set is denoted by D$d$Q$q$. We use the D3Q19 set, together with the Bhatnagar-Gross-Krook (BGK) collision operator, on grids with cubic cells. The implementation is based on the Palabos library \citep{palabos}.

The dimensions of the computational domain are ($6.3h \times 2h \times 2.1h$) for simulations of smooth wall channels presented below and ($6.3h \times (2h+l) \times 2.1h$) for the simulations with filaments.
 Grids with the different resolutions are denoted by  G1, G2 and G3, respectively, and are listed in table~\ref{tab:grids}.
 %
 The grid G1 is used for the results presented in sections \ref{sec:results} and \ref{sec:extendend_discussion}. For this grid, the resolution is $\delta x^+ = 2$ for $\Rey_\tau = 180$, giving a grid size of ($568 \times 181 \times 190$). However, the grid is refined with a factor of two at the upper and lower walls, up to $y^+ \approx 40$. The refinements are made with an overlap of one coarse grid spacing as described by \citet{lagrava12}. To minimize mass and momentum imbalances, a correction step is included before the collision step at the interface nodes, similar to the method described by \citet{kuwata16}, however adapted to overlapping nodes. As is commonly done in LBM simulations of channel flows, we use a constant applied pressure gradient to drive the flow, implemented with the Guo forcing scheme \citep{guo02}. At the walls, the wet-node regularized boundary condition is used \citep{latt08}.

\begin{figure}
  \centering
  \begin{subfigure}{0.45\textwidth}
      \centering
      \includegraphics[width=6cm]{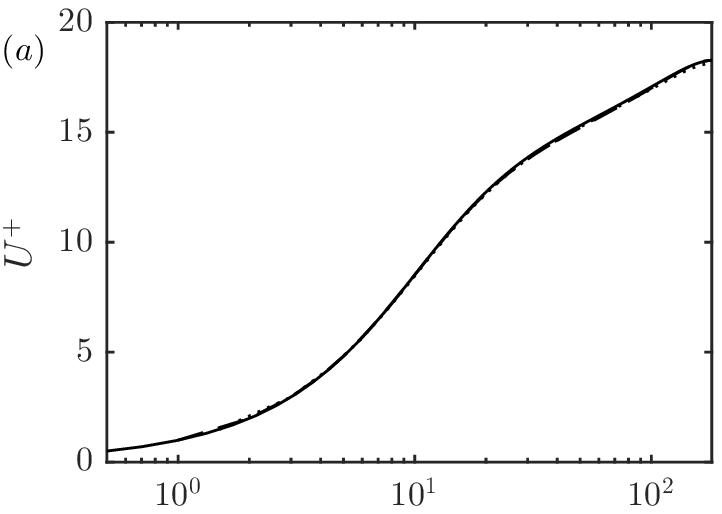}
  \end{subfigure}
  \begin{subfigure}{0.45\textwidth}
        \centering
        \includegraphics[width=6cm]{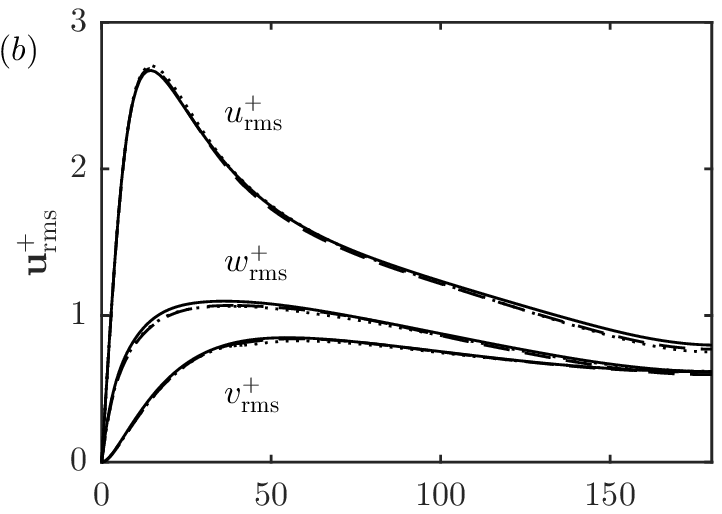}
  \end{subfigure}
  \begin{subfigure}{0.45\textwidth}
        \centering
        \includegraphics[width=6cm]{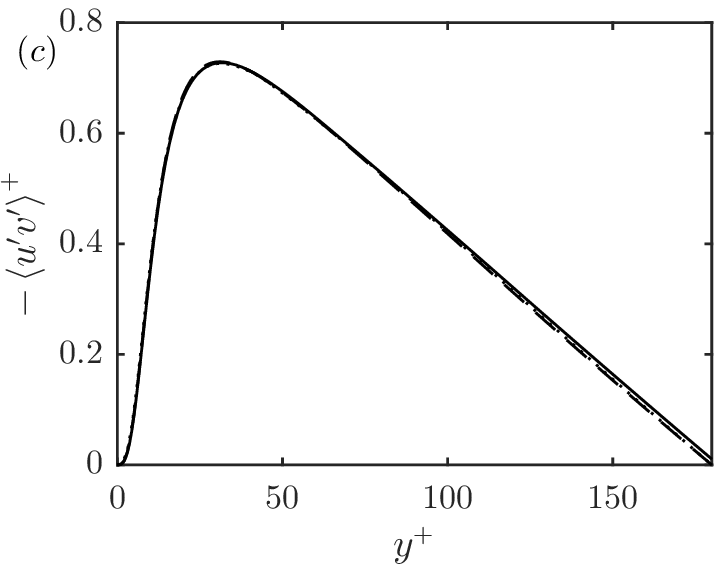}
  \end{subfigure}
  \begin{subfigure}{0.45\textwidth}
        \centering
        \includegraphics[width=6cm]{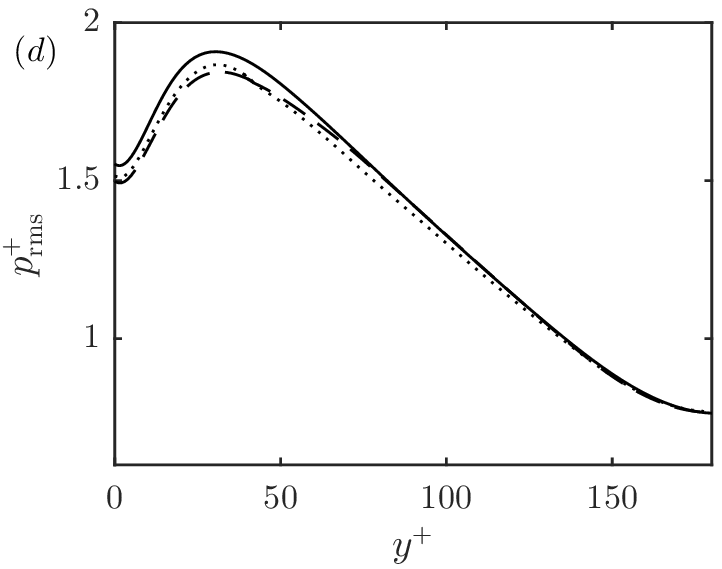}
  \end{subfigure}
  \vspace{3pt}
  \begin{tabular}{lll}
      \full~spectral & \longbroken~G1 & \dotted~G2
  \end{tabular}
  \caption{Statistics of a channel flow computed using the LBM employed in this work and computed using a spectral scheme \citep{lee15}. The ($a$) mean velocity, ($b$) velocity fluctuations, $(c)$ Reynolds shear stress and $(d)$ pressure fluctuations are shown. The agreement is considered satisfactory; errors are within a few percent for the two grids G1 and G2 when compared to the spectral results.}
  \label{fig:stats_LB}
\end{figure}

To validate the fluid solver, we compare a fully developed channel flow with smooth walls to the spectral DNS by \citet{lee15}. Two grids are used: G1 and G2, listed in tab.~\ref{tab:grids}. Due to the symmetry of the problem, the results from the two sides can be averaged. Fig.~\ref{fig:stats_LB} compares the mean velocity, r.m.s.~velocities, Reynolds shear stress (RSS) and r.m.s.~pressure to the spectral results.

Close to the wall, the r.m.s.~values of velocity and pressure fluctuations of the two grids G1 and G2 agree within $2\%$, indicating grid convergence. Comparing the r.m.s.~velocities and the RSS to the spectral results, the largest differences are around $3\%$. For the pressure fluctuations, however, the difference at the wall is slightly larger, around $3.6\%$. In the spectral DNS, the pressure is merely a product of the velocity field, whereas in the LBM it is a result of the density fluctuations. This may therefore result in a local violation of mass conservation at the wall, possibly originating from the boundary condition. Although there exist more sophisticated boundary conditions \citep{dorschner15}, we regard the current level of accuracy acceptable for a time scale analysis.

\subsection{Solid solver}
\label{sec:solid_solver}

The fluid-solid interaction is described by an immersed boundary approach, known as the external boundary force method (EBFM) \citep{wu10_particles}. This method uses a force to enforce the no-slip and the impermeability condition,  modelling the surface of a solid object. Surfaces of solid objects are discretised with a Lagrangian grid. This method has been used earlier to simulate fiber suspensions, both flexible and rigid, by \citet{wu10} and \citet{do-quang14}; the current implementation is based on the one by \citet{do-quang14}. The conversion of quantities between the Eulerian grid of the fluid and the Lagrangian grid of the filaments is performed with a discretisation of the Dirac delta function \citep{peskin02}.

\begin{figure}
  \centerline{\includegraphics[width=6cm]{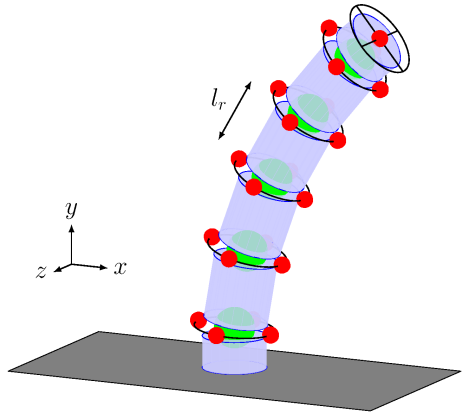}}
  \caption{(Colour online) Schematic illustration of the Lagrangian grid of a
    filament. Hinges (green) are connected by rods (blue), and each hinge has a ring of Lagrangian nodes (red). There are also additional nodes on the tip of the filament.}
  \label{fig:fiber_grid}
\end{figure}

In the current investigation, the dynamic Euler-Bernoulli equation \eqref{eq:Euler-Bernoulli} is discretised in a rod-hinge fashion, described below. The structure of the grid is shown in fig.~\ref{fig:fiber_grid}. This model was introduced by \citet{schmid00} and developed further by \citet{lindstrom07}, \citet{wu10} and \citet{do-quang14}. A similar model, using chains of spheres, was introduced by \citet{yamamoto93} and a model using chains of spheroids was introduced by \citet{ross97}. These models can be used both to describe flexible and rigid fibers.

A filament has $N_h$ hinges, with $N_n$ Lagrangian grid nodes in a ring around each hinge, together with $N_n^\text{lid}$ additional nodes on the lid. At the hinges, the filaments can deflect. The hinges of a filament are connected by rods and these can be extended or compressed. Hence, the filaments are extensible, however in practice the extensions and compressions of the rods are small, typically below $1\%$.
The direction of the rod between hinge $i-1$ and $i$ is parallel to the tangent unit vector
\begin{equation}
  \myvector{p}_i = \frac{\myvector{x}_i - \myvector{x}_{i-1}}{\norm{\myvector{x}_i - \myvector{x}_{i-1}}},
\end{equation}
where $\myvector{x}_{i}$ is the location of hinge $i$ and $\norm{\cdot}$ denotes the Euclidean norm. The first rod is assumed to be fixed to the wall, so that $\myvector{p}_1 = (0, 1, 0)^\top$, and for the last one we impose $\myvector{p}_{N_h+1} = \myvector{p}_{N_h}$.

In the Euler-Bernoulli equation, the bending moment of a beam, $M$, is assumed to be proportional to the curvature, $\kappa$, and the stiffness, giving $M = EI\kappa$, where $\kappa = \partial^2 q/\partial y^2$. For the rod-hinge model, the bending moment across hinge $i$ can be evaluated as
\begin{equation}
  \mathbf{M}_i = EI\frac{\arccos{(\myvector{p}_{i+1} \cdot \myvector{p}_i)}}{l_\mathrm{r}}\frac{\myvector{p}_{i+1} \times \myvector{p}_i}{\norm{\myvector{p}_{i+1} \times \myvector{p}_i}},
\end{equation}
where the first fraction is an estimation of the local curvature and the last fraction gives the direction. Considering an infinitesimal beam element, there is a local balance between the bending moment and the torque that the shear force, $S$, gives rise to, giving $S = \partial M /\partial y$. In our discrete model, however, we are not interested in the shear force, but the resulting force on a discrete segment of length $l_\mathrm{r}$, corresponding to the length of a rod or one hinge. This force, $F$, is given by the change in the shear force over a length $l_\mathrm{r}$. Using a linear approximation, $F = l_\mathrm{r}\partial S/\partial y = l_\mathrm{r} \partial^2 M /\partial y^2$. It can be noted that with the estimation of the curvature as the second derivative of the displacement, $F = l_\mathrm{r} EI \partial^4 q/\partial y^4$, which corresponds to the first term in the Euler-Bernoulli equation \eqref{eq:Euler-Bernoulli}. We approximate this force by central differences, taking into account the direction of the rods,
\begin{equation}
  \myvector{F}_i^\mathrm{h} = \frac{\mathbf{M}_{i-1} \times \myvector{p}_i}{l_\mathrm{r}} + \frac{\mathbf{M}_{i+1} \times \myvector{p}_{i+1}}{l_\mathrm{r}} - \frac{\mathbf{M}_{i} \times (\myvector{p}_{i+1} + \myvector{p}_{i})}{l_\mathrm{r}}.
\end{equation}
For the end hinge, the terms including $\myvector{p}_{i+1}$ are removed.

\begin{table}
  \centering
  \begin{tabular}{ll}
    Predicted $\tilde{q}/a$\hspace{0.5cm}     & Measured $\tilde{q}/a$       \\  [3pt]
    0.1                    & 0.093                  \\
    0.5                    & 0.489                  \\
    1.0                    & 0.918                  \\
    3.0                    & 2.579
  \end{tabular}
  \caption{Predicted static displacement, computed using the static Euler-Bernoulli equation, and measured static deflection using the considered rod-hinge model. The measured static deflection was approximated as the displacement of the top hinge.}
  \label{tab:static_validation}
\end{table}

When a rod is compressed or extended by a fractional change in length $\epsilon$, it results in a stress given by Hook's law, $\sigma = E\epsilon$. The resulting force on a hinge, corresponding to this stress, is
\begin{equation}
  \myvector{F}_i^\mathrm{r} = -EA\frac{\norm{\myvector{x}_i - \myvector{x}_{i-1}} - l_\mathrm{r}}{l_\mathrm{r}}\myvector{p}_i + \\ EA\frac{\norm{\myvector{x}_{i+1} - \myvector{x}_{i}} - l_\mathrm{r}}{l_\mathrm{r}}\myvector{p}_{i+1}
\end{equation}
For the end hinge only the first term remains.

The force from the fluid, $\myvector{F}_i^\mathrm{fluid}$, is calculated by summing the fluid forces on the ring of Lagrangian nodes, given by the EBFM. For the top hinge, the forces on the additional nodes of the lid are included. The ring of nodes at each hinge is tilted so that its normal is $(\myvector{p}_{i+1} + \myvector{p}_i)/\norm{\myvector{p}_{i+1} + \myvector{p}_i}$.
The total force on a hinge is then $\myvector{F}_i = \myvector{F}_i^\mathrm{h} + \myvector{F}_i^\mathrm{r} + \myvector{F}_i^\mathrm{fluid}$
, and the boundary condition implies that the total force on the first hinge is zero, $\myvector{F}_1 = 0$. With the explicit expression for the force on the hinges, the acceleration of each hinge can be calculated and they can be advected with the corresponding velocity.

The filament model has been validated in the static limit by applying a tip force at the top hinge and comparing it to the analytical prediction from the Euler-Bernoulli equation \eqref{eq:Euler-Bernoulli}. This was done for the geometrical parameters in tab.~\ref{tab:filament_cases} and four different deflection amplitudes. The results are summarised in tab.~\ref{tab:static_validation}, with errors around $9\%$ for smaller deflections ($\tilde{q} \lesssim a$), measuring the deflection of the top hinge. The deviation is slightly higher for the case of $\tilde{q} = 3a$ ($14\%$).
However  the analytical prediction --  in contrast to the rod-hinge model -- neglects the contribution from the first derivative of the displacement in the curvature and does not correct for the displacement of the beam in the vertical direction \citep{bisshopp45}. It therefore loses accuracy for deflections comparable to the length of the filament ($l = 10a$). Accounting for this, the error is reduced to $7\%$. The error can be further reduced to around 1\%, if accounting for the small deviation between the top hinge and the actual tip of the filament (see fig.~\ref{fig:fiber_grid}). We thus find the hinge-rod model as a reasonable numerical description of Euler-Bernoulli-type of filaments.
We characterized the dynamic response of a filament by measuring the natural frequency in the step-response of an applied tip force. For both cases I and III, the difference between the measured and the predicted frequency (by eq.~\ref{eq:exact_natural_frequency}) was less than $1\%$.

A grid refinement study was also  performed to validate the fluid-solid interaction under turbulent conditions. For this study, we use the grids G1 and G3, having $\delta x^+ \approx 1$ and $\delta x^+ \approx 0.75$ near the filamentous wall, respectively (see tab.~\ref{tab:grids}).
For case III, the drag and the r.m.s.~velocities were increased by around 5-6\% using G3.
When it comes to  identifying different fluid-surface interaction regimes by varying the time scale ratio $T/T_\mathrm{f}$, we find the accuracy provided by G1 acceptable. 





\section{Comparison of  slow and fast filamentous beds}
\label{sec:results}

\begin{figure}
    \centering
    \begin{subfigure}{0.45\textwidth}
        \includegraphics[width=6cm]{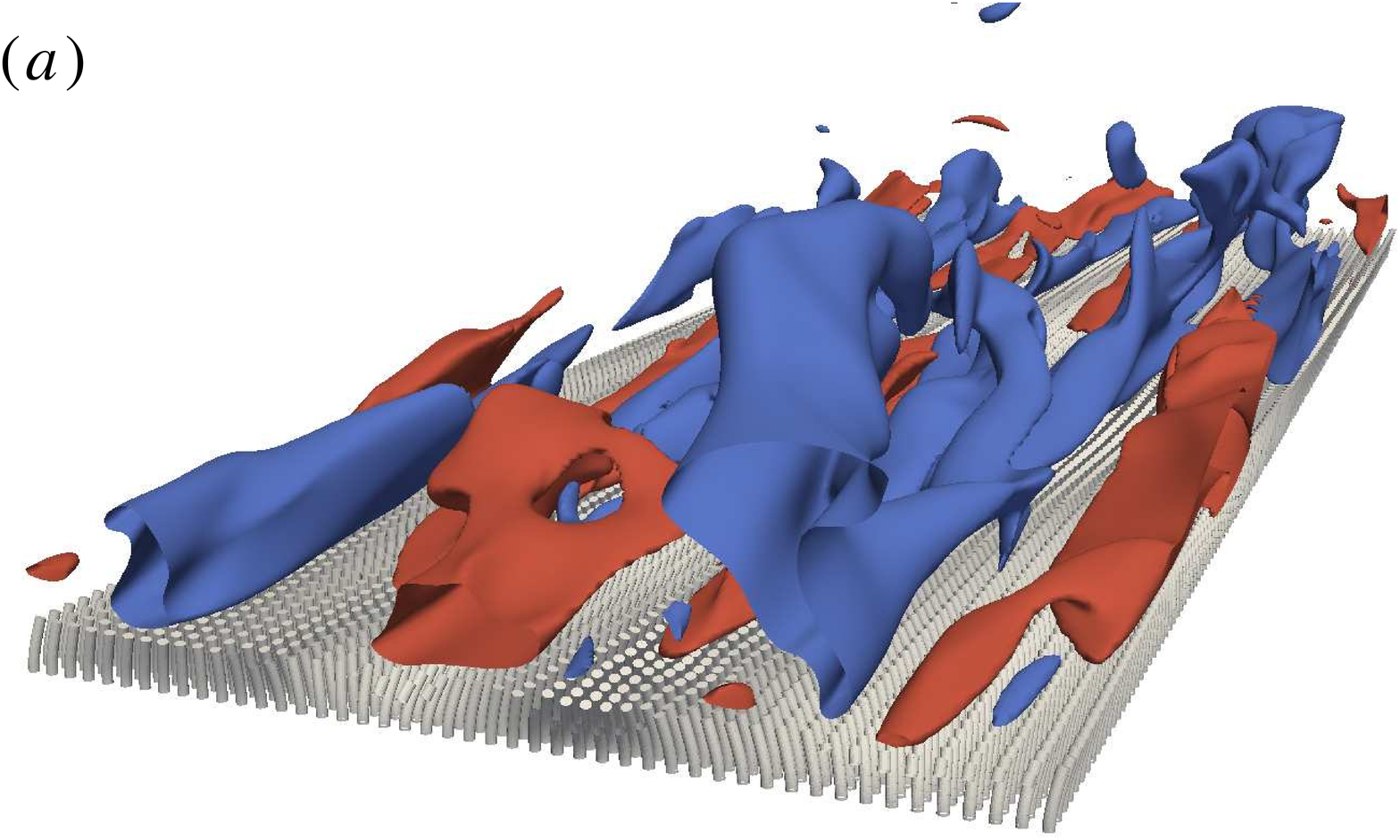}
    \end{subfigure}
    \begin{subfigure}{0.45\textwidth}
        \includegraphics[width=6cm]{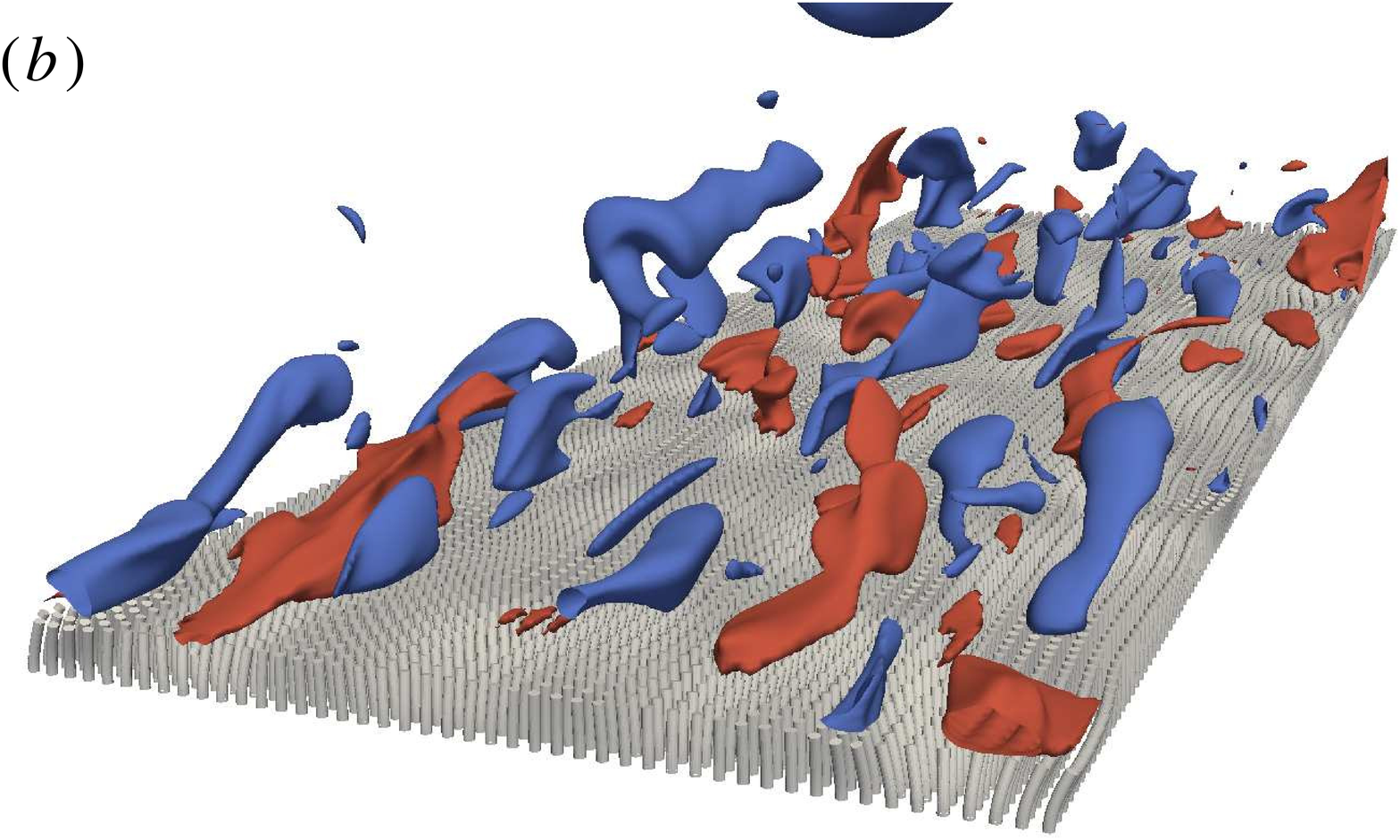}
        \vspace{0.5cm}
    \end{subfigure}
    \caption{(Colour online) Isosurfaces of the streamwise velocity fluctuations  $u'^+ = \pm 3$ for case ($a$) I (heavy bed) and ($b$) III (light bed). Filaments are shown with gray color. The mean flow is directed into the page. The case with the higher filament resonance frequency, III, has a higher isotropy in the velocity field and streaks are absent.}
    \label{fig:isotropy_visualization}
\end{figure}

In this section,
we show that the modification of the turbulent velocity field to a high extent depends on the resonance frequency of the filaments. Flow over low resonance frequency beds behave similarly to flow over a smooth wall channel, whereas the configurations with high resonance frequency indicate an absence of streaks and fluctuation fields of higher isotropy. This can be observed by the isosurfaces of the velocity fields of cases I and III (heavy and light filaments, respectively), shown in fig.~\ref{fig:isotropy_visualization}. These two configurations are discussed in detail in sections \ref{sec:filament_movement} and \ref{sec:turbulence_characteristics}, comparing filament movement and turbulence behaviour respectively.

\begin{figure}
    \begin{minipage}{8cm}
        \centering
        \begin{subfigure}{1.0\textwidth}
            \centering
            \includegraphics[width=8cm]{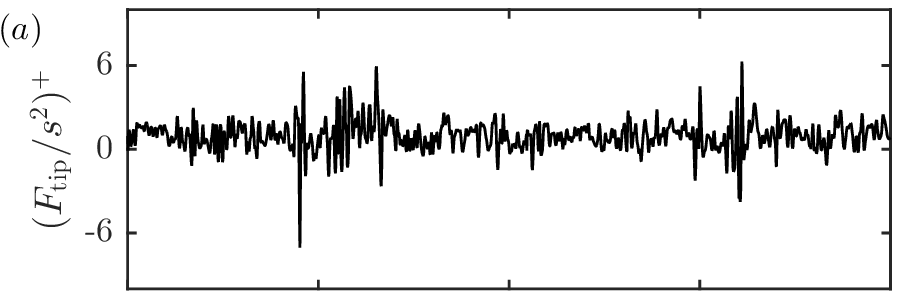}
        \end{subfigure}
        \begin{subfigure}{1.0\textwidth}
            \centering
            \includegraphics[width=8cm]{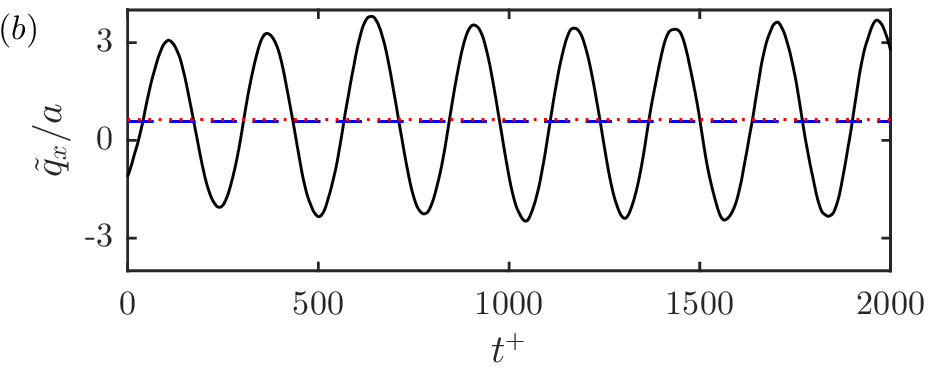}
        \end{subfigure}
    \end{minipage}
    \begin{minipage}{6cm}
        \begin{subfigure}{1.0\textwidth}
            \centering
            \includegraphics[width=6cm]{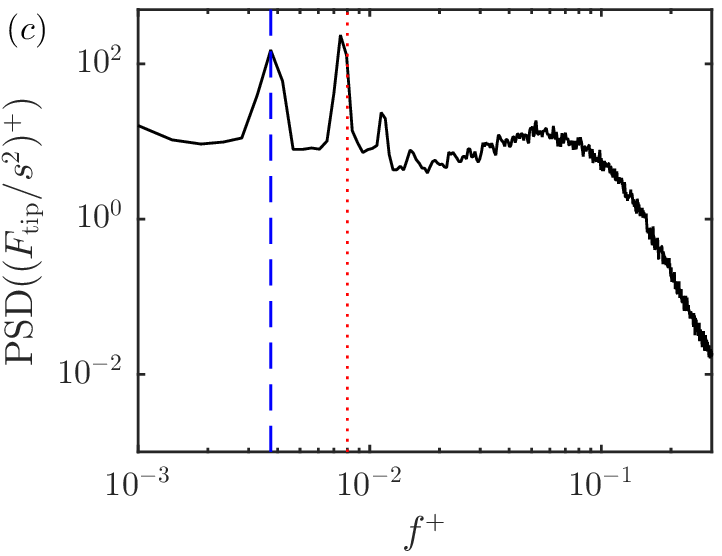}
        \end{subfigure}
    \end{minipage}
    \begin{minipage}{8cm}
        \centering
        \begin{subfigure}{1.0\textwidth}
            \centering
            \includegraphics[width=8cm]{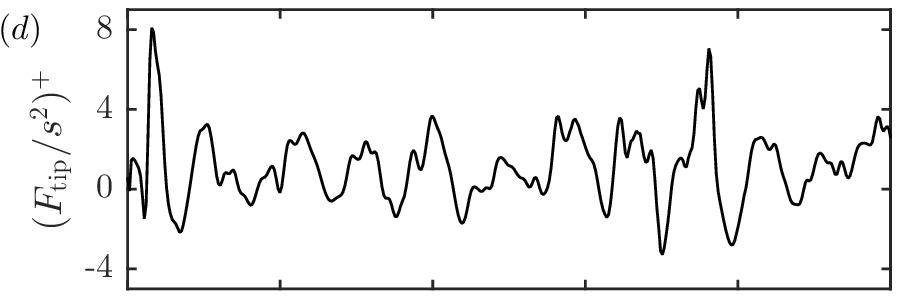}
        \end{subfigure}
        \begin{subfigure}{1.0\textwidth}
            \centering
            \includegraphics[width=8cm]{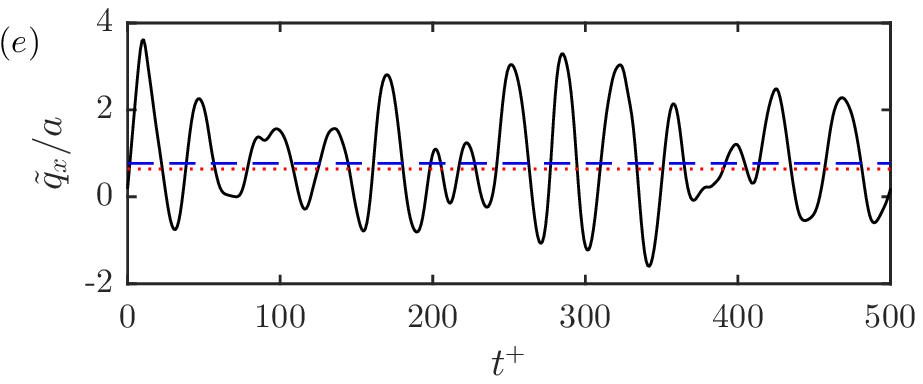}
        \end{subfigure}
    \end{minipage}
    \begin{minipage}{6cm}
        \begin{subfigure}{1.0\textwidth}
            \centering
            \includegraphics[width=6cm]{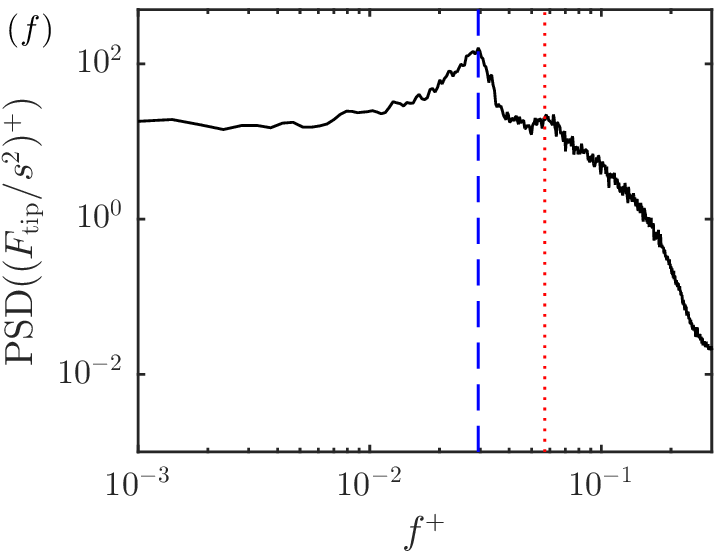}
        \end{subfigure}
    \end{minipage}
    \caption{(Colour online) Time series of ($a$) the forcing, ($b$) the tip displacement of a single filament and ($c$) the PSD of the force for case I (heavy bed). Corresponding data for case III (light bed) are shown in ($d$)-($f$). The measured (\longbroken) and predicted (\dotted) mean displacement, $Q$, are shown in ($b$) and ($e$). The measured (\longbroken) and predicted (\dotted) natural frequency, $f_\mathrm{n}$, are shown in ($c$) and ($f$).
    When the filaments move, they also generate a force, and this results in a peak in the PSD at the natural frequency and smaller peaks at multiples of the resonance frequency.
    %
    %
    }
    \label{fig:time_series}
\end{figure}

When results are reported in wall-units, they are based on the shear stress of the specific configuration. For the wall with filaments, the friction velocity is
\begin{equation}
    u_\tau^\mathrm{f} = \sqrt{\tau_\mathrm{wall}^\mathrm{f}/\rho},
    \label{eq:u_tau_f}
\end{equation}
where
\begin{equation}
    \tau_\mathrm{wall}^\mathrm{f} = \mean{F_\mathrm{tip}}/s^2
    \label{eq:wall_shear_stress}
\end{equation}
represents the effective total shear stress at the plane $y=0$.
From the solid-fluid interaction scheme (section \ref{sec:solid_solver}), the force from the fluid on a filament, $\mathbf{F}^\mathrm{fluid}$, is known. Further, we neglect the wall-shear stress of the wall to which the filaments are attached. To calculate the mean of the tip force, we can then consider the streamwise momentum balance of a filament cell (fig.~\ref{fig:geometry}$b$),
\begin{equation}
    \mean{F_\mathrm{tip}} = \mean{F_x^{\mathrm{fluid}}} + l s^2\left.\frac{\dif p}{\dif x}\right|_\mathrm{applied}.
\end{equation}


\subsection{Bed response}
\label{sec:filament_movement}


For the heavy bed (case I), the force on one filament is shown in fig.~\ref{fig:time_series}$a$, the displacement of the same filament in \ref{fig:time_series}$b$ and the power spectral density (PSD) of the force in \ref{fig:time_series}$c$. We employ standard spectral analysis for the estimation of the spectra \citep{brandt11}, however a short summary is provided in appendix \ref{sec:transfer_function}.
From the time series, the strong low pass filtering property of the heavy filamentous bed is apparent: the force in fig.~\ref{fig:time_series}$a$ contains a large range of frequencies, while the filament movement in fig.~\ref{fig:time_series}$b$ is dominated by the resonance frequency and lower frequencies.
The mean streamwise displacement, $Q = 0.58a$ (blue dashed line in \ref{fig:time_series}$b$), is close to the displacement (red dotted line) estimated from expression \eqref{eq:mean_disp}.
Around the mean, we observe from \ref{fig:time_series}$b$ nearly periodic fluctuations of the filaments with a time period of $T^+_f\approx 300$ which corresponds to $f^+=0.004$ (dashed vertical line in \ref{fig:time_series}$c$).
This can be compared to the resonance frequency estimated from \eqref{eq:exact_natural_frequency}, $f^+=0.0080$ (red dotted vertical line in \ref{fig:time_series}$c$). It is thus clear that the slow response of the bed is dominated by the low resonance frequency of the filaments.
%
We note that the tip force occasionally attains a negative value due to the interaction between adjacent filaments.

Corresponding figures for case III are shown in fig.~\ref{fig:time_series}$d$--$f$. This bed is lighter and thus also much faster, which can clearly be observed from the high correlation between the signals in fig.~\ref{fig:time_series}$d$ and \ref{fig:time_series}$e$. The correlation coefficient \citep{freund10} between the displacement and the tip force in the streamwise direction was found to be $r_\mathrm{III} = 0.71$, while for I it was $r_\mathrm{I} = 0.24$. The lower correlation coefficient of I reflects its slower and more resonant filaments. The dominant measured frequency of the forcing (dashed vertical line in fig.~\ref{fig:time_series}$f$) is $f^+ \approx 0.03$, whereas the frequency estimated from \eqref{eq:exact_natural_frequency} is $f^+=0.057$ (red dotted vertical line in fig.~\ref{fig:time_series}$f$).


\begin{table}
    \centering
    \begin{tabular}{lllll}
        Case\hspace{0.5cm} & $\Rey_\tau^\mathrm{f}$\hspace{0.2cm} & $\Delta D$\hspace{0.5cm} & $\tilde{q}_{x, \mathrm{rms}}/a$ \hspace{0.2cm}& $\tilde{q}_{z, \mathrm{rms}}/a$ \\  [3pt] 
        A   & 183 & 0.02 & ---  & --- \\
        B   & 204 & 0.76 & ---  & --- \\
        I   & 186 & 0.06 & 2.61 & 1.39 \\
        II  & 198 & 0.39 & 2.10 & 2.53 \\
        III & 200 & 0.48 & 0.90 & 0.87 \\
        IV  & 183 & 0.03 & 1.21 & 0.35 \\
        V   & 197 & 0.34 & 1.71 & 2.27 \\
        VI  & 198 & 0.37 & 0.80 & 0.81 \\
    \end{tabular}
    \caption{The friction Reynolds number of the wall with filaments, $\Rey_\tau^\mathrm{f}$, the global drag increase, $\Delta D$, and r.m.s.~values of the filament displacements obtained from numerical simulations of cases A,B,I-VI.}
    \label{tab:results}
\end{table}

\begin{figure}
     \centering
     \begin{subfigure}{0.45\textwidth}
         \centering
         \includegraphics[width=6cm]{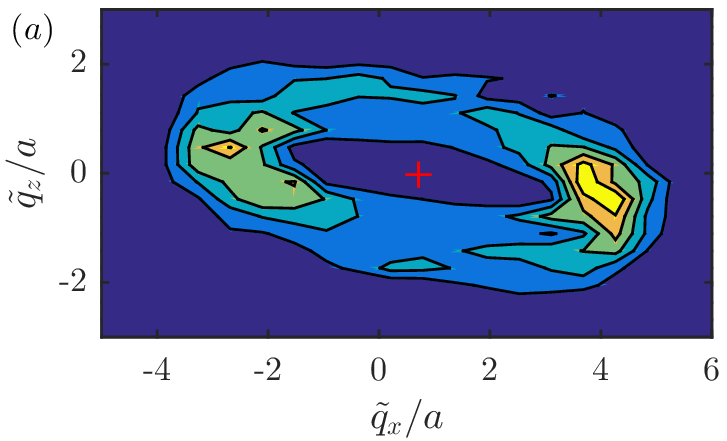}
     \end{subfigure}
     \begin{subfigure}{0.45\textwidth}
         \centering
         \includegraphics[width=6cm]{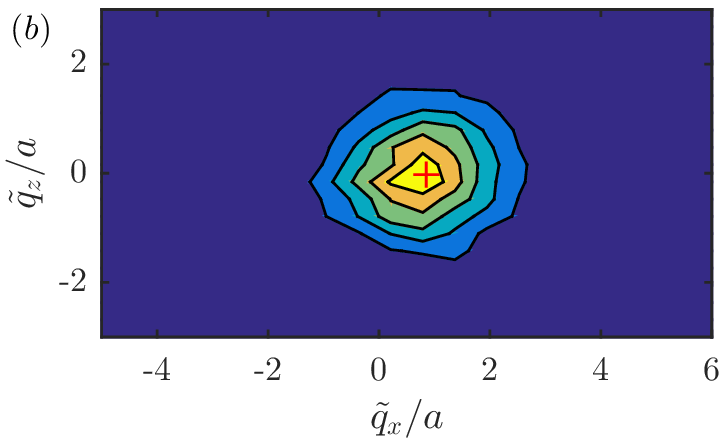}
     \end{subfigure}
     \caption{(Colour online) Contours of the probability density function of the filament tip displacement, $\tilde{\mathbf{q}}$, in the $xz$-plane, for case ($a$) I and ($b$) III. Areas of high probability density are yellow, whereas areas of low probability density are blue. The mean displacement is  marked with a red cross. The slight asymmetry of the probability density of I in the spanwise direction is attributed to the finite size of the sample.
     }
     \label{fig:q_dist}
 \end{figure}

The r.m.s.~values of the streamwise and spanwise displacements are
\begin{equation}
    \begin{array}{ll}
        (\tilde{q}_{x}, \tilde{q}_{z})_\mathrm{rms} = (2.61, 1.39)a &\text{ for I and } \\
        (\tilde{q}_{x}, \tilde{q}_{z})_\mathrm{rms} = (0.90, 0.87)a &\text{ for III,}
     \end{array}
\end{equation}
as reported in tab.~\ref{tab:results}. For case I, the streamwise r.m.s.~value is almost twice the spanwise value, whereas for III they are similar. The small difference of the two components for III indicates a high isotropy of the forcing and thus of the fluid velocity fluctuations.
%
\begin{figure}
    \centering
    \includegraphics[width=13cm]{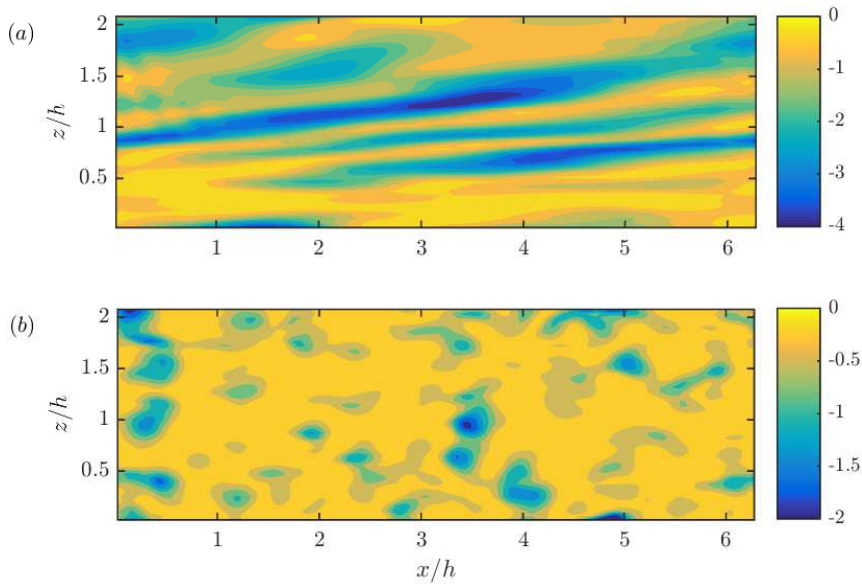}

    \caption{(Colour online) Wall-normal filament deformation, $\tilde{q}_y/a$, i.e.~at the plane $y=0$, for ($a$) case I and ($b$) case III, at one time instant.
    For case I, low frequency streaky structures are apparent, whereas the deformation field of case III is more isotropic.}
    \label{fig:bed_deformation}
\end{figure}
This difference can be understood in more detail by comparing the wall-normal displacement field for case I and III, respectively, at one time instant (fig.~\ref{fig:bed_deformation}). The wall-normal displacement is a consequence of the spanwise and streamwise displacement, since in practice, the filaments are inextensible; it thus provides a measure of the total displacement. The displacement field of I is dominated by large streamwise structures, similar to streaks, while for III, there are no such structures and the field is relatively isotropic. Further statistical information of the filament motion is provided in fig.~\ref{fig:q_dist}, showing contours of the probability density function (PDF) of the displacement in the $xz$-plane for I (\ref{fig:q_dist}a) and III (\ref{fig:q_dist}b). As indicated by the displacement r.m.s.~values, the displacements of the filaments in I are much larger in the streamwise direction than the spanwise direction. For III, the distribution is similar in the stream- and spanwise directions.

 From fig.~\ref{fig:bed_deformation}, it can be observed that the displacement field of I has a streak-like structure, whereas III has a much more isotropic displacement field. For the slow and heavy bed, adjacent filaments move in phase, creating the streamwise elongated streaky structures. Generally, for a smooth surface, the spanwise spacing between streaks is approximately $\Delta z^+ = 100$, corresponding to $\Delta z = 0.5 h$, whereas the length of streaks is around $\Delta x^+ = 1000$, corresponding to $\Delta x = 5 h$ \citep{pope01}. This is similar to what is observed for case I in fig.~\ref{fig:bed_deformation}$a$. Also, for I, the time scale of the filament movement has the same order of magnitude as the passing of a streak: considering the speed of a high speed streak to be around $u^+ = 3$ (from fig.~\ref{fig:u_rms}$a$), the streak travels a distance $\Delta x^+ = 800$ during one period of oscillation of a filament at the resonance frequency. 

\begin{figure}
    \centering
    \begin{subfigure}{0.55\textwidth}
        \centering
        \includegraphics[width=6.5cm]{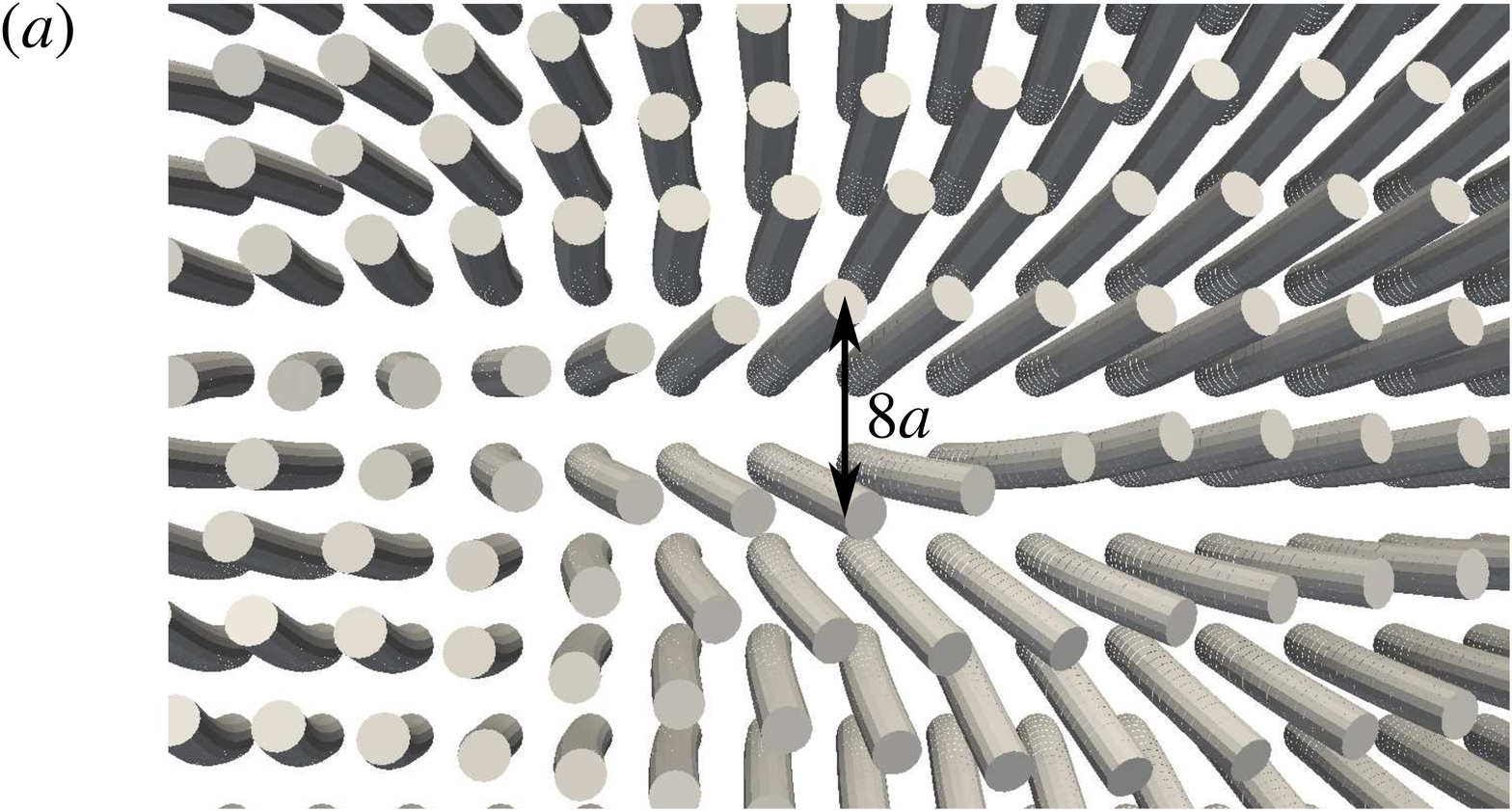}
    \end{subfigure}
    \begin{subfigure}{0.4\textwidth}
        \centering
        \includegraphics[width=6cm]{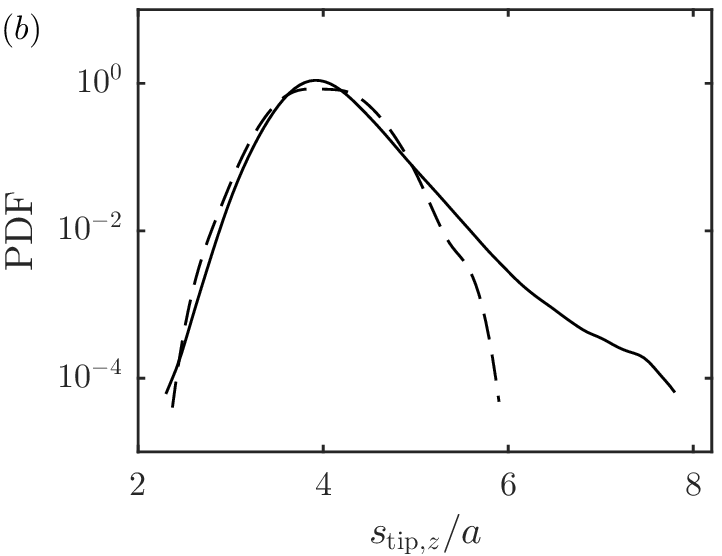}
    \end{subfigure}
    \caption{($a$) Zoomed-in top view of an event of large filament displacements for case III, with the direction of mean flow to the right. The center-to-center distances at the filament tips is locally enlarged, increasing the permeability. ($b$) The probability density function of the center-to-center distance in the spanwise direction, $s_{\mathrm{tip},z}$, for III (\full) and I (\longbroken).}
    \label{fig:large_disp}
\end{figure}

For case III, fig.~\ref{fig:bed_deformation}$b$ reveals localized regions of relatively large displacements, around $\Delta x = 0.5h$ in size. This size corresponds to $10s$, where $s$ is the center-to-center distance of the filaments.
These regions of separated filaments are created by fluctuations with negative wall-normal velocity moving towards the wall, similar to sweep events. A zoomed-in view of the filaments at such an event is shown in fig.~\ref{fig:large_disp}$a$. We observe that these events create center-to-center distances of the filament tips in the spanwise direction, $s_{\mathrm{tip},z}$, up to $8a$. Therefore, these patches have a higher permeability, increasing locally the transport of mass and momentum from the free flow to the bed.  The PDF of $s_{\mathrm{tip},z}$ is presented in fig.~\ref{fig:large_disp}$b$. The mean is $\mean{s_{\mathrm{tip},z}} = s$, and $s_{\mathrm{tip},z}$ is constrained to be larger than one filament diameter, $2a$.
This can be compared to the corresponding PDF of case I, where we observe a tighter distribution around the mean $s=4a$.

We do not observe any clear evidence of  periodic vortex shedding due to filaments; there is no common peak in the spectra of the forcing for the different cases that could relate to such events (compare fig.~\ref{fig:time_series}c and f). Vortex separation also seems improbable from an estimation of the Reynolds number based on the filament diameter. 


\subsection{Turbulence modification}
\label{sec:turbulence_characteristics}
Having discussed the behaviour of the filament movement of cases I and III in section \ref{sec:filament_movement}, we here discuss the modification of turbulence over the two beds. In order to separate the contributions of a permeable surface from a deformable surface, we will also compare the turbulence modifications to that of the rigid but permeable configurations, A and B. Case A has the same geometry as I and III, whereas B has twice the filament center-to-center distance, namely $s = 8a$, and thus a higher permeability than A (the permeability scales with $s^2$).

\subsubsection{Drag characteristics}
A first quantification of the effects of the filaments on the turbulence can be done by measuring the change in drag. The local drag increase of the filament wall can be characterised by $\Rey_\tau^\mathrm{f} = u_\tau^\mathrm{f}h/\nu$.
From the tabulated data in tab.~\ref{tab:results}, we observe that the friction Reynolds number for the slow bed (case I) is very close to that of the smooth channel ($Re_\tau=180$), whereas for the fast bed (case III) we have $Re_\tau^\mathrm{f}=200$. Interestingly, the rigid configuration with higher permeability (case B) has a friction Reynolds number ($Re_\tau^\mathrm{f}=204$) very close to case III, giving a first indication that B and III modify the near wall turbulence in a similar way.

The drag increase of the complete channel is defined by
\begin{equation}
    \Delta D = \frac{c_\mathrm{f}}{c_\mathrm{f}^\mathrm{0}} - 1,
\end{equation}
where $c^0_\mathrm{f}$ is friction coefficient of the smooth symmetric channel.
%
%
The friction coefficient of the channel with non-smooth wall is based on the stress of both the top and bottom walls, i.e.
\begin{equation}
    c_\mathrm{f} = \frac{\frac{1}{2}(\tau_\mathrm{wall}^\mathrm{f}+\tau_\mathrm{wall}^\mathrm{t})}{\frac{1}{2}\rho U_\mathrm{b}^2}.
\end{equation}
Here, $\tau_\mathrm{wall}^\mathrm{f}$ is the effective total shear stress of the bottom wall (eq.~\ref{eq:wall_shear_stress}) and $\tau_\mathrm{wall}^\mathrm{t}$ is the shear stress of the top wall. Since a constant pressure gradient is used, the numerator is constant and $\Delta D$ is produced by change in the bulk velocity,
\begin{equation}
    U_\mathrm{b} = \frac{1}{2h}\int_0^{2h} U \dif y.
\end{equation}
%
%
%
Tab.~\ref{tab:results} reports $\Delta D$ for the  cases I, III, A and B. We observe a similar trend as for the local friction Reynolds number;
the drag increase is much lower for case I than III, with $\Delta D = 0.06$ and $\Delta D = 0.48$ respectively; the drag of I is similar to A ($\Delta D = 0.02$), whereas case III has a drag of the same order as that of B ($\Delta D = 0.76$).

The fast flexible bed (case III) thus increases drag by an amount comparable to that of the rigid -- but highly permeable  -- surface (case B). Indeed, from the PDF of $s_{\mathrm{tip},z}$ in fig.~\ref{fig:large_disp}$b$, it is found that III locally may have a permeability close to that of B, $s \approx 8a$. As the permeability is increased, the fluctuation of the wall-normal velocity is also increased (as will be shown later). It has been found that for rough walls, increased wall-normal velocity fluctuations is related to an increase in drag \citep{orlandi03, orlandi06}. Also for porous walls there is a correlation between wall-normal velocity fluctuations and skin-friction drag, which can be attributed to Kelvin-Helmholtz vortices \citep{breugem06}. We did however not observe such large-scale spanwise rollers for case III.

Apart from the permeability, elasticity can also be a source for drag increase \citep{kim14}. The  drag increase of soft compliant walls is often attributed to quasi-two-dimensional waves of the surface, propagating in the streamwise direction. These waves have been observed for canopy-flows, then termed monami, caused by large-scale Kelvin-Helmholtz vortices \citep{nepf12}. For the parameters  considered here, no such waves have been observed, and the displacement fields displayed in fig.~\ref{fig:bed_deformation} are absent of the characteristic bands such waves create.

%

\begin{figure}
    \centering
    \makebox[\textwidth] {
    \begin{subfigure}{0.45\textwidth}
        \centering
        \includegraphics[width=6cm]{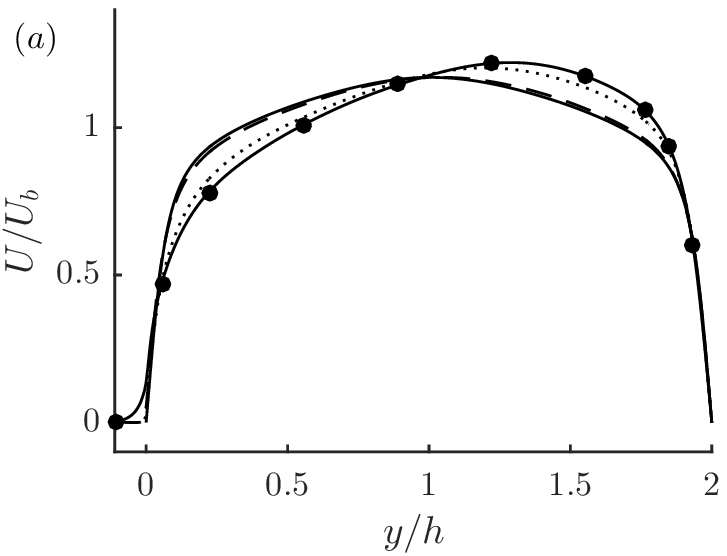}
    \end{subfigure}
    \begin{subfigure}{0.45\textwidth}
        \centering
        \includegraphics[width=6cm]{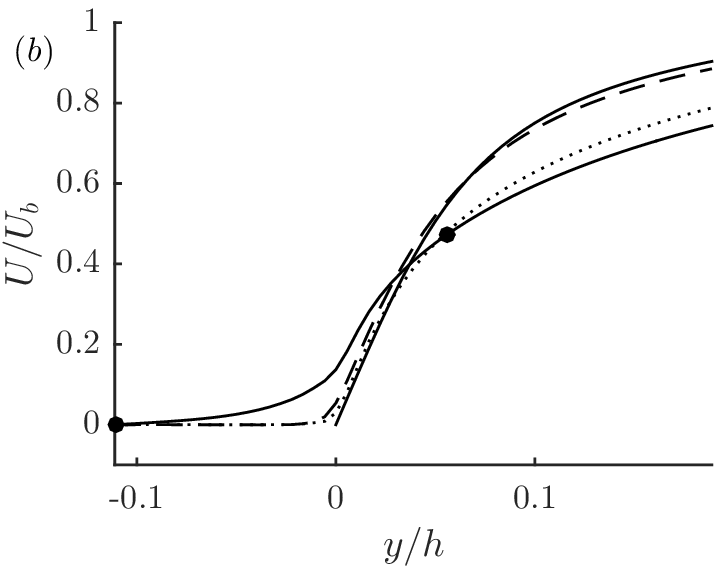}
    \end{subfigure}
    } \\ [3pt]
    \begin{tabular}{llll}
        \full~smooth & ---$\bullet$---~case B (rigid) & \longbroken~case I (heavy) & \dotted~case III (light)
    \end{tabular}
    \caption{Mean velocity profiles in outer units, showing ($a$) the complete channel and ($b$) the region closest to the lower wall. A shift of the profile to the right indicates an increase in wall drag.}
    \label{fig:u_outer}
\end{figure}

The drag increase observed for case III is  due to the local increase in permeability
and the mechanism  has similarities to the drag increase induced by rigid wall roughness \citep{orlandi03, orlandi06}.
The major effect of roughness is a decrease of the mean velocity profile compared to a smooth wall for the same friction Reynolds number, $\Rey_\tau$. 

Mean velocity profiles of cases I, III and B, together with the smooth wall case, are presented in fig.~\ref{fig:u_outer}$a$, with a zoomed-in view in fig.~\ref{fig:u_outer}$b$. Note that outer scaling is used, in order to characterise the modification of the velocity profile in the complete channel.
We observe only a slight difference between the heavy bed (I) and the symmetric smooth wall profile.
In contrast, for the light bed (III),
the profile is skewed towards the upper wall (without filaments). The mean profile of the rigid but sparse bed (B) shows a similar shift towards the upper, smooth wall. This indicates a high drag increase at the lower wall of these cases.

The mean velocity inside filamentous bed ($y<0$), driven by the external pressure gradient, is much larger for case B than the other configurations (fig.~\ref{fig:u_outer}$b$), due to the higher filament center-to-center distance. 


\begin{figure}
    \centering
    \makebox[\textwidth] {
    \begin{subfigure}{0.45\textwidth}
        \centering
        \includegraphics[width=6cm]{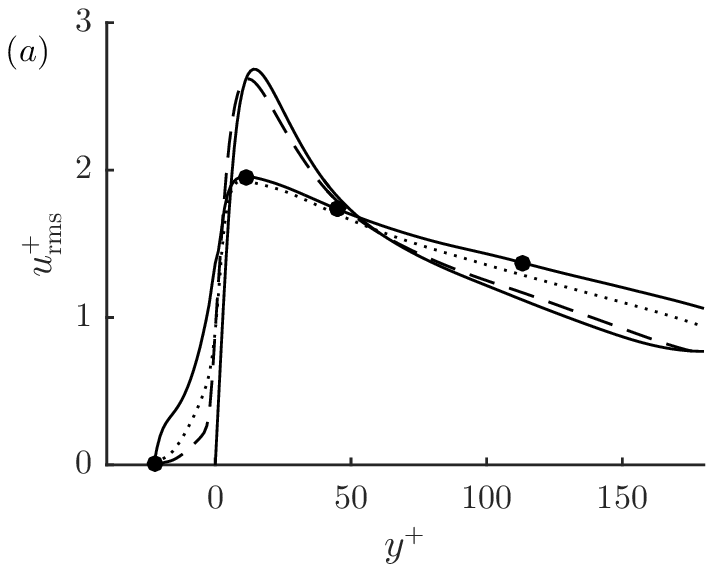}
    \end{subfigure}
    \begin{subfigure}{0.45\textwidth}
        \centering
        \includegraphics[width=6cm]{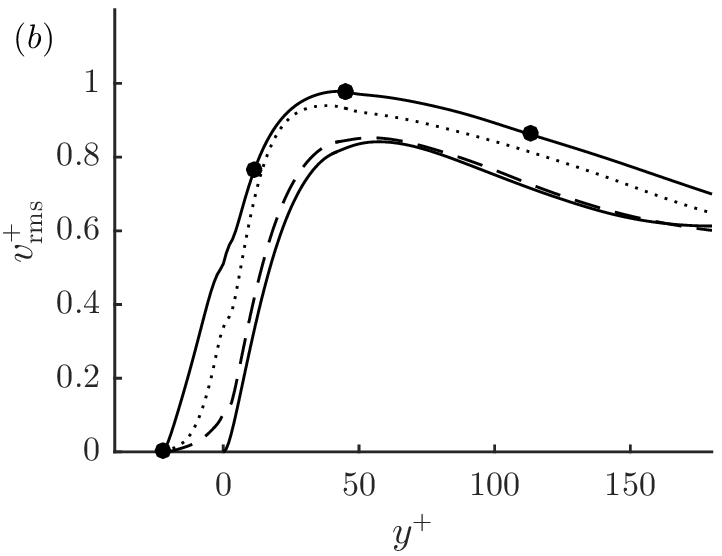}
    \end{subfigure}
    } \makebox[\textwidth] {
    \begin{subfigure}{0.45\textwidth}
        \centering
        \includegraphics[width=6cm]{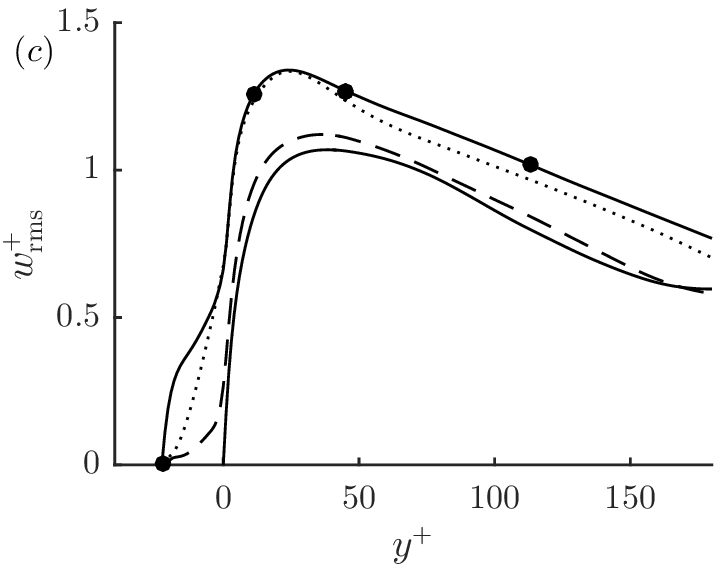}
    \end{subfigure}
    \begin{subfigure}{0.45\textwidth}
        \centering
        \includegraphics[width=6cm]{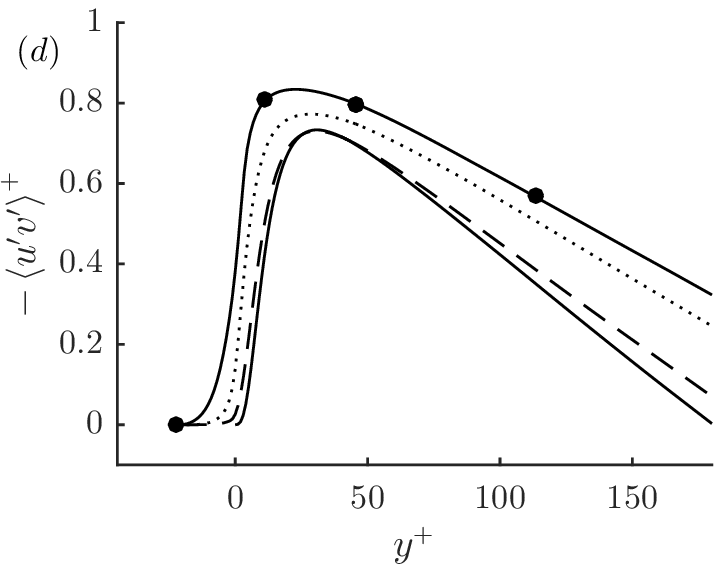}
    \end{subfigure}
    } \\ [3pt]
    \begin{tabular}{llll}
        \full~smooth & ---$\bullet$---~case B (rigid) & \longbroken~case I (heavy) & \dotted~case III (light)
    \end{tabular}
    \caption{Profiles of ($a$) streamwise ($b$) wall-normal and ($c$) spanwise r.m.s.~velocity, together with the Reynolds shear stress ($d$). Scaling is based on $u_\tau^\mathrm{f}$. The filaments of Case B and III result in an increased isotropy by decreasing the streamwise component, while increasing the wall-normal and spanwise components.}
    \label{fig:u_rms}
\end{figure}

\subsubsection{Turbulent fluctuations}
\begin{figure}
    \centering
    \includegraphics[width=13cm]{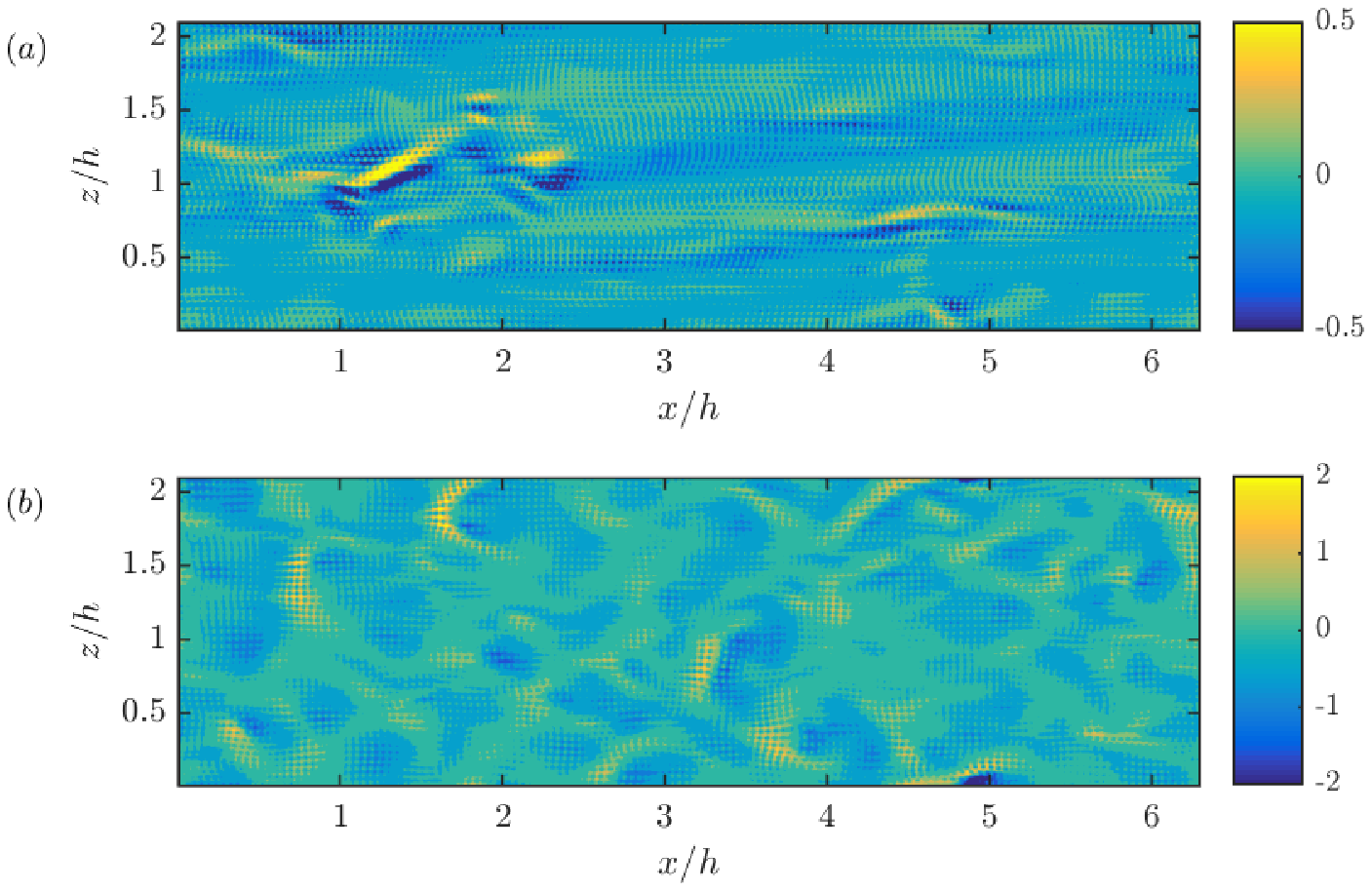}

    \caption{(Colour online) Wall-normal velocity fluctuations, $v^+$, at the crest plane of the filaments, $y=0$, for ($a$) case I and ($b$) case III, at one instant. Scaling is based on $u_\tau^\mathrm{f}$. The velocity field of I contains relatively isolated sweep events, whereas strong velocity fluctuations are much more frequent for III.}
    \label{fig:wallnormal_fluctuations}
\end{figure}

In order to understand in more detail how the flow inside the bed interacts with turbulence just above the bed, we present the turbulent fluctuations scaled with local values of the friction velocity, $u_\tau^\mathrm{f}$.
R.m.s.~velocities and the Reynolds shear stress (RSS) at the filament wall are presented in fig.~\ref{fig:u_rms}.
%
Let us first characterise the fluctuations above the bed ($y>0$). For all three normal components and the RSS, case I (dashed lines) shows only small deviations from the smooth wall profiles (solid lines).
We can  conclude that case I is not only slow, making it act as a rigid rough wall, but it is also so dense that  it acts as a smooth wall. The turbulence-surface interaction of case I is thus essentially one-way coupled, i.e. surface deforms slowly due to streaks (fig.~\ref{fig:bed_deformation}$a$), but the flow is essentially left unmodified by the surface.

In contrast, for case III, one observes that above the bed, the streamwise velocity component (dotted line in \ref{fig:u_rms}$a$) is prominently reduced, whereas the wall-normal and spanwise components (figs.~\ref{fig:u_rms}$b$ and $c$) are increased. The rigid case B (filled circle symbols) behaves similarly.
%
The decrease of the peak of the streamwise component can be attributed to the reduction (or even absence) of streaks. The increase of the peak values of the wall-normal and spanwise components can be attributed to disturbances caused by the filamentous wall, in particular to ejections from the interior of the bed, similarly as observed for rough walls by \citet{orlandi06}. The wall-normal velocity field at one instant is shown in fig.~\ref{fig:wallnormal_fluctuations} for I and III at the crest plane of the filaments, $y = 0$. The character of the velocity fields reflect those of the displacement fields, fig.~\ref{fig:bed_deformation}, with a high isotropy for III.

The strong interaction with the fast filamentous wall of case III is related to a reduction of the so-called wall-blocking effect, commonly observed for  highly permeable walls \citep{breugem06}. Wall-blocking occurs as fluid moving towards the wall cannot penetrate the wall and must change direction to move parallel to a wall, creating a "splat" event \citep{perot95}. Energy is transferred from the wall-normal component to the tangential components, increasing tangential turbulence intensity.

For case III, which shows local regions of higher permeability, the fluid moving towards the wall penetrates the filamentous bed. This can be observed in fig.~\ref{fig:u_rms}$b$, where $v^+_\mathrm{rms}$ is relatively large for $y<0$. As a consequence, velocity fluctuations are transported into the bed, inducing larger $u^+_\mathrm{rms}$ and $w^+_\mathrm{rms}$ between the filaments. The RSS, fig.~\ref{fig:u_rms}$d$, confirms this by also having large values for $y<0$. Note that the r.m.s.~wall-parallel displacements, $(\tilde{q}_x,\tilde{q}_z)_\mathrm{rms}$, are smaller for the fast bed (III) compared to the slow bed (case I). The velocity fluctuations inside the fast bed show the reverse trend: velocity r.m.s.~are larger in case III than case I. This indicates that the fluctuations inside the interior of the bed, are not primarily due to movement of the filaments, but rather due to penetration and ejections of turbulent fluctuations (similar to as rough wall).

\section{Transfer function analysis}
\label{sec:extendend_discussion}


Case I and III represent two separate phenomena. The filaments of I $(T>T_\mathrm{f})$ are too slow to adapt to quick changes of the turbulence. They respond only to the slowly moving streaks, and do not disturb the overlying turbulent flow, and hence can be said to have a one-way coupling to the turbulence. On the other hand, the filaments of III $(T<T_\mathrm{f})$ capture more of the turbulent time scales and act like a filament bed with higher permeability (case B).
Here we will characterise the additional filamentous beds presented in tab.~\ref{tab:filament_cases} and compare their response to turbulence using a simple model.
%

\subsection{A lumped spring model of the surface}
\label{sec:spring_model}
In this section, we analyse the filament time scale more quantitatively by formulating a transfer function representing the filaments. With a quasi-static assumption, it is possible to form a transfer function using the Euler-Bernoulli equation \citep{brucker07}. The spatial shape of the filaments is then assumed to be the same in the static and the dynamic case. The resulting equation describes a damped harmonic oscillator, and the model can therefore be seen as a lumped spring model of the filament dynamics. Analytical results are presented below, but the detailed derivation is provided in appendix~\ref{sec:transfer_function_derivation}.

The transfer function in the streamwise direction, giving the tip deflection amplitude $\hat{q}_x(\omega)$ from the tip force $\hat{F}_\mathrm{tip}(\omega)$, as function of the angular frequency $\omega$, is
\begin{equation}
    |H(\omega)| = \frac{1}{\sqrt{\left(1 - \left(\frac{\omega}{\omega_\mathrm{n}^\mathrm{LSM}}\right)^2\right)^2 + 4\zeta^2\left(\frac{\omega}{\omega_\mathrm{n}^\mathrm{LSM}}\right)^2}},
    \label{eq:transfer_function}
\end{equation}
normalised with the mean displacement
\begin{equation}
    \mean{\tilde{q}} = \frac{1}{3}\frac{\mean{F_\mathrm{tip}}l^3}{EI}
    \label{eq:mean_disp}
\end{equation}
so that $|H(0)| = 1$.
In the lumped spring model (LSM), the natural frequency is
\begin{equation}
    \omega_\mathrm{n}^\mathrm{LSM} = 2\pi f_\mathrm{n}^\mathrm{LSM} = \frac{2}{l^2}\sqrt{\frac{2EI}{A\rho_\mathrm{s} + \chi}},
    \label{eq:resonance_frequency}
\end{equation}
and the damping ratio is
\begin{equation}
    \zeta = \mean{\tilde{q}}\omega_\mathrm{n}^\mathrm{LSM}\left[\frac{c}{2\mean{F_\mathrm{tip}}}\right ],
    \label{eq:damping}
\end{equation}
where $\mean{F_\mathrm{tip}} = \hat{F}_\mathrm{tip}(0)$ is the mean force on a filament. Here, $c$ is a damping constant, defined in appendix \ref{sec:transfer_function_derivation}, that depends only on the geometry. Eq.~\eqref{eq:mean_disp}, together with \eqref{eq:cauchy_number}, show that $Q^* = 3\mean{\tilde{q}}/a$, and eq.~\eqref{eq:resonance_frequency} resemble the natural frequency, \eqref{eq:exact_natural_frequency}, but with a slightly different coefficient.

The fraction in the brackets in expression \eqref{eq:damping} depends, to a first approximation, only on the geometry, so that for a given bed $\zeta$ scales with the product of the resonance frequency and the mean amplitude, $\mean{\tilde{q}}\omega_\mathrm{n}^\mathrm{LSM}$. This is reasonable since $\mean{\tilde{q}}\omega_\mathrm{n}^\mathrm{LSM}$ is a characteristic speed of the filaments. When inertia dominates over viscous damping ($\zeta < 1/\sqrt{2}$), $|H|$ is fairly constant for $f < f_\mathrm{n}^\mathrm{LSM}$, with a slight bump at, or close to, $f_\mathrm{n}^\mathrm{LSM}$, after which it decreases. In this sense, it corresponds to a low pass filter, apparent from fig.~\ref{fig:time_series}$d$ and fig.~\ref{fig:time_series}$e$. A similar transfer function is present in the spanwise direction; the filament geometry is isotropic except for the reconfiguration induced by the mean displacement.

\begin{figure}
    \centering
    \begin{subfigure}{0.3\textwidth}
        \centering
        \includegraphics[width=4cm]{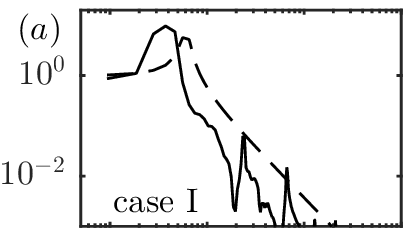}
    \end{subfigure}
    \begin{subfigure}{0.3\textwidth}
          \centering
        \includegraphics[width=4cm]{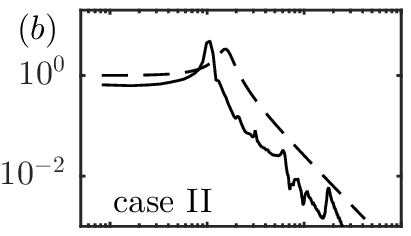}
    \end{subfigure}
    \begin{subfigure}{0.3\textwidth}
          \centering
        \includegraphics[width=4cm]{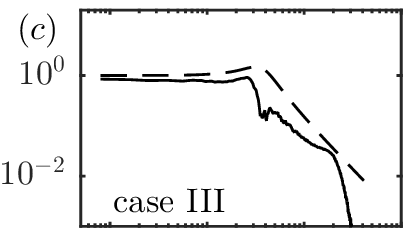}
    \end{subfigure}
    \begin{subfigure}{0.3\textwidth}
          \centering
        \includegraphics[width=4cm]{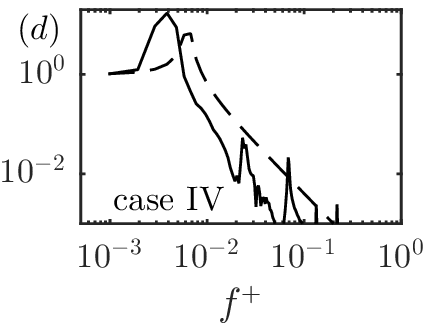}
    \end{subfigure}
    \begin{subfigure}{0.3\textwidth}
          \centering
        \includegraphics[width=4cm]{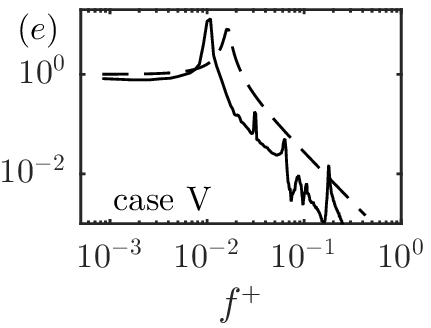}
    \end{subfigure}
    \begin{subfigure}{0.3\textwidth}
          \centering
        \includegraphics[width=4cm]{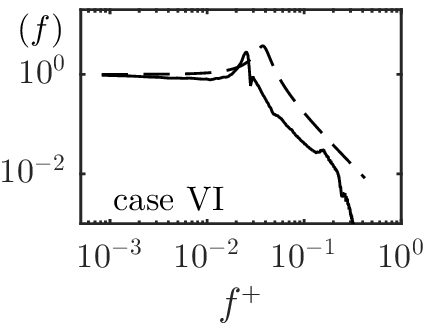}
    \end{subfigure}  \\ [3pt]
    \begin{tabular}{ll}
        \full~measured & \quad \longbroken~predicted
    \end{tabular}
    \caption{Absolute value of measured transfer functions between $F_\mathrm{tip}$ and $\tilde{q}_x$ for case I--VI in ($a$)--($f$), respectively. Predicted transfer functions are also shown, calculated by eq.~\eqref{eq:transfer_function}.}
    \label{fig:H}
\end{figure}

Note that there exist other surface time scales; if instead inertia is neglected, $A\rho_\mathrm{s} + \chi$ is assumed to small and hence the natural frequency, $\omega_\mathrm{n}^\mathrm{LSM}$, is large. A new time scale then determines the response in eq.~(\ref{eq:transfer_function}),
\begin{equation}
    T = 4\pi\frac{\zeta}{\omega_\mathrm{n}^\mathrm{LSM}} = 2 \pi \frac{cl^3}{6EI},
\end{equation}
which is independent of the density of the filaments. This is the time scale in the viscous regime, determining the cut-off frequency of the filaments when damping dominates. For example, this is the case of more sparsely placed sensor filaments \citep{brucker07}. It is also similar to the poroelastic time scale discussed by \citet{skotheim04, skotheim05lub}. The latter time scale determines how fast the pressure (and thus the flow) inside a poroelastic medium equilibrate to the far field conditions. The speed of which the surface equilibrates to the surrounding is thus set by fluid transport, rather than by the deformation of the filaments.

\subsection{Beds with higher resonance peak or smaller mean displacements}
Next, we will compare the analytical predictions of the transfer functions from previous subsection with the transfer functions computed from numerical simulations (appendix~\ref{sec:transfer_function}).

In addition to cases I and III, studied in depth in section \ref{sec:results}, the top row of fig.~\ref{fig:H}$a$--$c$, also shows the transfer function for case II.  Configuration II has filaments with resonance frequency ($f_\mathrm{n}^+=0.024$) in between I and III, and one may expect that the filaments will yield to forces of higher frequencies than in case I ($f_\mathrm{n}^+=0.0080$), but not as high as in III ($f_\mathrm{n}^+=0.057$).
By comparing figs.~\ref{fig:H}$b$ and $c$, we observe that case II has a significantly higher resonance peak than III. This means that displacements at this frequency are higher for the same imposed force. This is reflected by the r.m.s~values of the displacements, which are significantly larger for II than III,
%
\begin{equation}
    \begin{array}{ll}
        (\tilde{q}_{x}, \tilde{q}_{z})_\mathrm{rms} = (2.10, 2.53)a &\text{ for II and } \\
        (\tilde{q}_{x}, \tilde{q}_{z})_\mathrm{rms} = (0.90, 0.87)a &\text{ for III}.
     \end{array}
\end{equation}
As a result of the resonant behavior, above the bed ($y>0$), the velocity fluctuations in all three directions are enhanced. In particular, we observed that the peak value of the spanwise r.m.s~velocity, $w_\mathrm{rms}^+$, is larger by 20\% and the wall-normal, $v_\mathrm{rms}^+$, by 5\%.
However, higher resonance of the surface does not result in a larger entrainment into the bed. At the filaments tips $(y=0)$,  $v_\mathrm{rms}^+$ is lower by $31\%$ (comparing II to III); this indicates that even though case II has significantly larger filament displacements than case III, it corresponds to a surface of lower apparent permeability than III. It thus seems that due to the high resonance peak, the filaments respond stronger to the turbulent flow. However,  the lower resonance frequency and the higher resonance peak of the surface compared to III inhibit the surface to comply to sweep events to the same extent.


The corresponding analytical predictions of the transfer functions of the filaments (obtained from eq.~\ref{eq:transfer_function}) are shown with a dashed line in fig.~\ref{fig:H}$a$--$c$ for case I, II and III, with added mass according to equation \eqref{eq:added_mass}. The parameter $c$ in eq.~(\ref{eq:damping}) is fixed to one value ($c=5.9\mu l$) for all the transfer functions shown in fig.~\ref{fig:H}, chosen by fitting the data. This value gives damping ratios in the interval $0.02 \le \zeta \le 0.4$.
We observe that the model predicts the response behavior of the surface reasonably well. The model is empirical in the sense that it requires  estimates of lumped parameters that depend on the fluid and flow properties, i.e. added mass $\chi$ and damping coefficient $c$.

We performed three additional simulations (with same fixed $c$ and $\chi$ as before), namely cases IV-VI in tab.~\ref{tab:filament_cases}. Compared to cases I-III, these configurations have a lower predicted mean displacement but the same predicted resonance frequencies. In this way, we can assess whether the fluid-surface interaction for heavier but stiffer beds also is determined by $f_\mathrm{n}^+$ alone. The transfer function of these configurations are shown in fig.~\ref{fig:H}$d$--$f$ and tab.~\ref{tab:results} reports the drag and the r.m.s.~displacements. We note that these configurations behave very similarly to I--III;  (i) The case with lowest natural frequency, IV, has a drag close to that of A ($\Delta D = 0.03$ and $\Delta D = 0.02$ respectively), whereas the drag is much higher for V and VI ($\Delta D = 0.34$ and $\Delta D = 0.37$); (ii) Case IV also has a lower isotropy of the flow velocity than V and VI, apparent from the displacement r.m.s.~values.

This parametric study demonstrates that both the model and the simulations show a consistent behavior when it comes to bed response to turbulent forcing. In other words, the fluid-surface interaction is indeed determined by resonance frequency only, and one may either change the mass of the filaments or their elasticity to tune the response.

 From fig.~\ref{fig:H}, we observe for all cases that the frequency of the resonance peak is over-predicted by the model. This is also observed in fig.~\ref{fig:time_series}c and f. Looking at the expression for the location of the natural frequency, eq.~\eqref{eq:resonance_frequency}, this appears to be related to an underprediction of $\chi$. Possibly, the modelling of the filament-filament coupling by $\chi = \rho s^2$ (eq.~\ref{eq:added_mass}) can be improved, although this very simple model suffices to capture the  fluid-structure interaction and to provide physical insight.

\section{Conclusions}
\label{sec:conclusion}
We have presented a surface time-scale analysis using numerical simulations of a bed of filaments in a turbulent channel flow.
By keeping the geometry of the filaments  fixed, but changing the filament density, $\rho_\mathrm{s}$, and Young's modulus, $E$, we could systematically investigate how a filamentous bed interacts with turbulent flow for different characteristic surface time scales $T$. 

In particular, if $T\gg T_\mathrm{f}$, where $T_\mathrm{f}$ is a characteristic time scale of the turbulent forcing, the bed will not respond to turbulent fluctuations above the bed and the surface can then be described as a rigid rough surface. If $T\gtrsim T_\mathrm{f}$, the surface may capture some of the slowly evolving turbulent structures, such as streaks, with minimal modification of the turbulent flow.  Such a surface may be useful for sensor design, where one may obtain information of large-scale turbulent structures near the wall.

On the other hand, if $T\ll T_\mathrm{f}$, the bed will equilibrate instantaneously with the turbulent fluctuations, and therefore interact with the turbulent flow as if it was a rigid surface with higher (non-uniform) permeability. These surfaces disturb the turbulence significantly, making it more isotropic and increasing the drag. However, they also increase the entrainment of free fluid into the bed. The exchange of mass and momentum is not fueled by shear-layer instabilities (such as Kelvin-Helmholtz instabilities), but because the surface can quickly comply with sweep/ejection events, opening up its interface to the free flow, and thus allowing for an increased flux of fluctuations into the bed. Such a surface may be useful in application where mixing and entrainment are beneficial.

We could also observe that when $T\sim T_\mathrm{f}$, the surface may resonate with the turbulence forcing, increasing significantly the displacement and velocity fluctuations. However, these surfaces also increase the wall-blocking behavior of the surface; the elastic solid does not have time to relax, and the surface behaves as material with ``memory'', and is therefore not able to instantaneously comply with the forcing induced by turbulence.

In this paper, we have focused on a particular configuration, namely a semi-dense  carpet in a turbulent channel flow at $Re_\tau\approx 180$. Below, we discuss a few limitations and  extensions of our work.
%
First, for the relatively dense hairy surfaces investigate here, the deformation of the filaments are primarily induced by the overlying turbulent shear forces, and not by pressure forces within the poroelastic medium.  In practice, this means that in order to obtain  $Q^*$ and $T^*$ of $\mathcal{O}(1)$, one needs dense, soft, high-aspect ratio filaments and a flow with a strong shear. For example, a turbulent channel flow (either water or air) at $Re_\tau\sim 4000$ with height $h\sim1000a$  and filaments with stiffness $B\sim 10^{-9}$ Nm$^2$ and length of $l\sim 100a$ would fall within this regime.
Direct numerical simulations of such a configuration is currently computationally too expensive; our numerical simulations are thus made feasible by reducing the friction Reynolds number as well as the size and the flexural rigidity of the filaments in order to preserve $Q^*$ and $T^*$ of $\mathcal{O}(1)$.

Second, the largest displacement amplitudes recorded in our simulations are around $3a$. Although these events are rather rare, real materials might experience non-linear effects, so that the Euler-Bernoulli equation is no longer applicable. In practical situations, the filaments might also be effected by movements of the surface to which they are attached. Especially for heavy filaments, surface vibrations can induce oscillatory motion of the filaments. Hence, the anchoring of the filaments to the surface needs to be taken into consideration in specific configurations.

Finally, other time/length scale interaction mechanisms may dominate in other configurations, in particular, for very dilute or very dense filaments. 
For example, the poroelastic time scale is the appropriate measure when the response of the bed is related to the time it takes for the interstitial viscous fluid  inside the surface to settle.
One may also expect different fluid-surface interaction for significantly larger and softer filaments (e.g. vegetation), where additional fluid and structural instabilities  -- such as inflectional shear-layers or propagating surface waves -- appear. These instabilities evolve on larger macroscopic length-scales than the microscopic pore size (i.e. distance between the filaments) of the surface, and therefore exploit large-scale motions to increase mixing and entrainment. These surfaces are more conveniently described using effective continuum/homogenisation approaches \citep{gopinath11, lacis17}.

Nevertheless, within the regime of validity of the current analysis, one may use the insight provided here --  using both  numerical simulations as well as the simple lumped-spring model -- to design surfaces for different objectives. For example, hairy surfaces are efficient in resisting the accumulation of microorganisms on submerged surfaces \citep{wan13}. At the same, coating the hull of a ship with a hairy surface -- that is not designed properly -- is  likely to increase drag significantly. The optimal design of the hairy surface in this context will thus maximize the resistance to bio-fouling and minimize the induced drag increase.
Fluid-surface interaction analysis for larger range of geometries and higher Reynolds numbers will be needed in the future to provide engineers appropriate tools and guidelines for designing complex surfaces.




\acknowledgments
This work was supported by SSF, the Swedish Foundation for Strategic Research (Future Leaders Grant FFL15:0001).
Simulations were performed at the PDC Centre for High Performance Computing (PDC-HPC) and at the High Performance Computing Center North (HPC2N) on resources provided by the Swedish National Infrastructure of Computing (SNIC).

\appendix
\section{Derivation of the  transfer function \eqref{eq:transfer_function}}
\label{sec:transfer_function_derivation}
The pressure acting on the filaments can be considered to have two parts. One is the macroscopically varying pressure, in this case by the explicitly applied pressure difference. The other is due to the variations at pore scale, with average zero across a cell, $\Delta p_\mathrm{pore} = 0$. The explicitly applied pressure difference over a cell must hence, under static conditions and for an in the wall-normal direction infinite bed, equal the drag of a filament,
\begin{equation}
    f_\mathrm{body} = -s^2\left.\frac{\dif p}{\dif x}\right|_\mathrm{applied},
    \label{eq:zero_gradient_limit}
\end{equation}
where $x$ is the direction of the applied pressure gradient. With the height of the filaments assumed to be infinite, the velocity gradient in the wall-normal direction is zero. This approximately holds for dense filament beds \citep{nepf12}.

We now consider the tip force. For finite $s$, there is a global modification of the velocity field, and not only in the asymptotic way as for a lone filament. If the fluid velocity at the filament tips is approximated as zero, the filament bed corresponds to an impermeable wall. In reality, momentum diffuses, and there will be a region of high shear, on average, around the filament tips. As a first approximation however, the impermeable wall approximation can be used, implying a force of
\begin{equation}
    F_\mathrm{tip} = s^2\mu\left.\frac{\partial U}{\partial y}\right|_{y=0},
    \label{eq:grad_force}
\end{equation}
where the filament tips are located at $y=0$. In this expression, it is assumed that a filament gets all the momentum transferred to a cell. This assumption is consistent with the zero-gradient limit of eq.~\eqref{eq:zero_gradient_limit}. In this first approximation, the velocity at the filament tips are neglected, and hence damping due to filament movement is not described, nor is momentum transfer due to Reynolds shear stress.

The magnitude of the force contributions can now be compared. The body force distribution gives a total force of $F_\mathrm{body} = f_\mathrm{body}l$. For a channel flow, with channel half-height $h$ and approximately symmetric wall-shear stress $\tau_\mathrm{wall}$,
\begin{equation}
    -\left.\frac{\dif p}{\dif x}\right|_\mathrm{applied}2h = 2\tau_\mathrm{wall} \quad \implies \quad -\left.\frac{\dif p}{\dif x}\right|_\mathrm{applied} = \frac{\tau_\mathrm{wall}}{h}.
\end{equation}
Hence,
\begin{equation}
    \frac{F_\mathrm{body}}{F_\mathrm{tip}} = \frac{-ls^2\left.\frac{\dif p}{\dif x}\right|_\mathrm{applied}}{s^2\mu\left.\frac{\partial U}{\partial y}\right|_{y=l}} = \frac{l \tau_\mathrm{wall}/h}{\tau_\mathrm{wall}} = \frac{l}{h} \ll 1,
    \label{eq:force_comparison}
\end{equation}
so that the force on the body of a filament is much smaller than the force at the tip.

Henceforth, we use a coordinate system with $y = 0$ at the filament base and $y = l$ at the tip, for simplicity. Considering the Euler-Bernoulli equation \eqref{eq:Euler-Bernoulli} in the static case, the inertial forces are zero. According to this equation, together with the boundary conditions \eqref{eq:boundary_conditions}, it then holds that
\begin{multline}
    q = \frac{1}{3}\frac{l^3}{EI}F_\mathrm{tip}\frac{1}{2}\left[-\left(\frac{y}{l}\right)^3 + 3\left(\frac{y}{l}\right)^2\right] + \frac{1}{8}\frac{l^3}{EI}F_\mathrm{body}\frac{1}{3}\left[\left(\frac{y}{l}\right)^4 - 4\left(\frac{y}{l}\right)^3 + 6\left(\frac{y}{l}\right)^2\right] \\ = \frac{1}{k_\mathrm{tip}}F_\mathrm{tip}\Phi_\mathrm{tip} + \frac{1}{k_\mathrm{body}}F_\mathrm{body}\Phi_\mathrm{body}.
\end{multline}
The expressions in the square brackets together with the rightmost fraction of each term represent the spatial shape of the filament associated with each force distribution, $\Phi_\mathrm{tip}(y)$ and $\Phi_\mathrm{body}(y)$, normalised so that $\Phi_\mathrm{tip}(l) = \Phi_\mathrm{body}(l) = 1$. The difference in the order of magnitude between the terms is determined by the forces, so that according to eq.~\eqref{eq:force_comparison}, the second one can be neglected. Looking at the displacement of the tip point, $\tilde{q}$, it holds that $k\tilde{q} = F$, where $F = F_\mathrm{tip}$ and
\begin{equation}
    k = k_\mathrm{tip} = 3\frac{EI}{l^3},
\end{equation}
with spatial shape $\Phi = \Phi_\mathrm{tip}$.

In the dynamic case, the inertial forces are included. It is also possible to include a damping of the filament motion, as a function of the velocity of the filaments, by elaboration of the body force, eq.~\eqref{eq:zero_gradient_limit}: If the fluid velocity around the filaments is small, the Reynolds number based on the diameter of the filaments is small, and if $\Rey_d \lesssim 1$, the Stokes equations are approximately valid. This is true for a dense filament bed or for short filaments, since they then are contained in the viscous sublayer. Based on the linearity of the Stokes equations, the force of the body of the filaments can be assumed to be  $f_\mathrm{body} = \mu U C_\mathrm{d}$, where $\mu = \rho\nu$ is the dynamic viscosity, $U$ is a characteristic fluid velocity at the filaments and $C_\mathrm{d}$ is the dimensionless drag coefficient of the laminar flow regime. In total,
\begin{equation}
    EI\frac{\partial^4 q}{\partial y^4} + (\rho_\mathrm{s} A + \chi)\frac{\partial^2 q}{\partial t^2} = \mu C_\mathrm{d} \left(U - \frac{\partial q}{\partial t} \right) \approx -\mu C_\mathrm{d}\frac{\partial q}{\partial t},
    \label{eq:dynamic_beam}
\end{equation}
where the velocity difference between the filament and the characteristic velocity of the bed is used in the body force. From the static case, it was seen that the body force gives a negligible contribution, however the term containing the filament velocity is kept as a model for the damping. This is a term of major importance for sensor filaments \citep{brucker07}, but the damping has been seen to be less important for densely placed filaments. Proportionality of the force to $\partial q/\partial t$ and $C_\mathrm{d}$ is not strictly valid if several adjacent filaments move in phase, however it is kept as a model here. Using the quasi-static approximation, the spatial shape is assumed to be the same as for the static case, so that
\begin{equation}
    q(y,t) = \Phi(y)\tilde{q}(t).
\end{equation}
From the static Euler-Bernoulli equation and the force at the boundary, $\int_0^l EI\partial^4 q/\partial y^4 \dif y = F_\mathrm{tip} = k\tilde{q}$, so that integrating eq.~\eqref{eq:dynamic_beam} over a filament yields
\begin{equation}
    k\tilde{q} + c\frac{\partial \tilde{q}}{\partial t} + m\frac{\partial^2 \tilde{q}}{\partial t^2} = F_\mathrm{tip}(t),
    \label{eq:damped_pendulum}
\end{equation}
where
\begin{equation}
    m = (\rho_\mathrm{s} A + \chi)\int_0^l \Phi(y) \dif y \quad \text{and} \quad
    c = \int_0^l \mu C_\mathrm{d}\Phi \dif y = c(\phi_s, a/l, \Rey_d).
\end{equation}
Evaluation of the integral of the spatial shape results in
\begin{equation} 
    m = \frac{3}{8}l(\rho_\mathrm{s} A + \chi).
\end{equation}
For the damping constant, $c$, a constant value is used, fitting the data (however, same for all cases).

Equation \eqref{eq:damped_pendulum} can be solved in the frequency domain. Considering a tip force $\hat{F}_\mathrm{tip}(\omega)$ and a tip deflection amplitude $\hat{q}_x(\omega)$ in the streamwise direction, the solution to the equation, forming the transfer function, is
\begin{equation}
    |H(\omega)| = \frac{\hat{q}_x(\omega)}{\hat{F}_\mathrm{tip}(\omega)/k} = \frac{1}{\sqrt{\left(1 - \left(\frac{\omega}{\omega_\mathrm{n}^\mathrm{LSM}}\right)^2\right)^2 + 4\zeta^2\left(\frac{\omega}{\omega_\mathrm{n}^\mathrm{LSM}}\right)^2}},
    \label{eq:transfer_function_appendix}
\end{equation}
with a scaling $|H(0)| = 1$, where the natural frequency is
\begin{equation}
    \omega_\mathrm{n}^\mathrm{LSM} = 2\pi f_\mathrm{n}^\mathrm{LSM} = \sqrt{\frac{k}{m}} = \frac{2}{l^2}\sqrt{\frac{2EI}{A\rho_\mathrm{s} + \chi}},
    \label{eq:natural_frequency}
\end{equation}
and the damping is described by
\begin{equation}
    \zeta = \frac{c}{2\omega_\mathrm{n}^\mathrm{LSM}m} = \frac{c}{2\sqrt{km}} = \left[\mean{\tilde{q}} = \frac{\mean{F_\mathrm{tip}}}{k} \right] = \mean{\tilde{q}}\frac{k}{\mean{F_\mathrm{tip}}}\frac{c}{2\sqrt{km}} = \mean{\tilde{q}}\omega_\mathrm{n}^\mathrm{LSM}\frac{c}{2\mean{F_\mathrm{tip}}}.
\end{equation}

\section{Statistical estimation of transfer function}
\label{sec:transfer_function}
The crosscorrelation between quantities $x(t)$ and $y(t)$ is defined as
\begin{equation}
     R_{xy}(\tau) = \frac{1}{T} \int_0^T x(t)y(t + \tau)\dif t, \quad T \rightarrow \infty,
\end{equation}
and the autocorrelation is obtained for $y(t) = x(t)$. Taking the Fourier transform of the crosscorrelation, the cross power spectral density (CPSD) is
\begin{equation}
     S_{xy} = \int_{-\infty}^\infty R_{xy}(\tau)e^{-i2\pi f \tau}\dif \tau.
\end{equation}
Similarly, the power spectral density (PSD) is the Fourier transform of the autocorrelation. The inverse relation is
\begin{equation}
     R_{xx}(\tau) = \int_{-\infty}^\infty S_{xx}e^{i2\pi f \tau}\dif f \implies R_{xx}(0) = \int_{-\infty}^\infty S_{xx} \dif f.
\end{equation}
At $\tau = 0$ the integral of $S_{xx}$ is $R_{xx}(0)$, i.e.~the r.m.s.~value squared, since "power" in PSD refers to the square of the r.m.s. value. The transfer function between $x$ and $y$ is calculated as
\begin{equation}
 H_1 = \frac{S_{xy}}{S_{xx}} \quad \text{ or } \quad H_2 = \frac{S_{yy}}{S_{yx}} = \frac{S_{yy}}{S_{xy}^*}.
\end{equation}
Which of these to use depends on the character of the expected error. If the error does not correlate to the measured input, $x(t)$, then $H_1$ is optimal, but if there in as error in the measured input but not the measured output, $y(t)$, then $H_2$ is optimal. In the case of filaments, $x(t)$ should be chosen as the force on the filaments, for example $F_\mathrm{tip}$, and $y(t)$ should be chosen as the corresponding displacement, $\tilde{q}_x$. In the numerical evaluations, $H_1$ is used. Averaging was performed using Hanning windows.

\bibliographystyle{jfm} \bibliography{references}

\end{document}